\documentclass[letterpaper]{article}

\usepackage{fullpage}
\usepackage{hyperref}
\usepackage[ruled,linesnumbered]{algorithm2e}
\usepackage{graphicx}
\usepackage{amsmath,amssymb,amsfonts}
\usepackage{enumerate}
\usepackage{subfigure,color,multirow}
\usepackage{tikz}
\usepackage{multirow}
\usepackage{hhline}
\usepackage{pgfplots}
\usepackage{pgfplotstable}
\DeclareGraphicsExtensions{.eps,.pdf,.png,.jpg}

\title{A distributed-memory package for dense Hierarchically Semi-Separable matrix computations using randomization}
\author{Fran\c{c}ois-Henry Rouet\footnotemark[1] \and
  Xiaoye S. Li\footnotemark[1] \and
  Pieter Ghysels\footnotemark[1] \and
  Artem Napov\footnotemark[2]
}

\hypersetup{
colorlinks=true,
citecolor=black,
linkcolor=black,
urlcolor=black
}

\usetikzlibrary{patterns}

\newcommand{\mytab}[2]{\setlength{\tabcolsep}{0pt}\begin{tabular}{p{0.36\textwidth}l}#1&#2\end{tabular}}

\newcommand{\compressed}{\texttt{COMPRESSED}}
\newcommand{\partially}{\texttt{PARTIALLY\_COMPRESSED}}
\newcommand{\untouched}{\texttt{UNTOUCHED}}

\begin{document}

\maketitle

\renewcommand{\thefootnote}{\fnsymbol{footnote}}
\footnotetext[1]{Lawrence Berkeley National Laboratory, Berkeley, CA 94720, USA. (\texttt{\{fhrouet,xsli,pghysels\}@lbl.gov})}
\footnotetext[2]{Universit\'e Libre de Bruxelles, B-1050 Brussels, Belgium.}
\renewcommand{\thefootnote}{\arabic{footnote}}

\begin{abstract}
We present a distributed-memory library for computations with dense structured matrices. A matrix is considered structured if its off-diagonal blocks can be approximated by a rank-deficient matrix with low numerical rank. Here, we use Hierarchically Semi-Separable representations (HSS). Such matrices appear in many applications, e.g., finite element methods, boundary element methods, etc. Exploiting this structure allows for fast solution of linear systems and/or fast computation of matrix-vector products, which are the two main building blocks of matrix computations. The \emph{compression} algorithm that we use, that computes the HSS form of an input dense matrix, relies on randomized sampling with a novel adaptive sampling mechanism. We discuss the parallelization of this algorithm and also present the parallelization of structured matrix-vector product, structured factorization and solution routines. The efficiency of the approach is demonstrated on large problems from different academic and industrial applications, on up to 8,000 cores.

This work is part of a more global effort, the STRUMPACK (STRUctured Matrices PACKage) software package for computations with sparse and dense structured matrices. Hence, although useful on their own right, the routines also represent a step in the direction of a distributed-memory sparse solver.

\end{abstract}

\maketitle

\section{Introduction}

\subsection{Background}

Many applications involve dense matrix computations with \emph{structured} (or \emph{low-rank}, or \emph{data-sparse}) matrices, i.e., matrices that are \emph{compressible} in some sense. In some applications, these matrices are rank-deficient or nearly so and can be readily compressed exactly or approximately using such algorithms as SVD, CUR~\cite{mahoney2009cur}, or a rank-revealing factorization. In many applications, the matrix is not (nearly) singular, but contains low-rank blocks, typically the blocks away from the main diagonal. Such matrices appear in the boundary element methods and finite element methods~\cite{chandrasekaran2010numerical,hackbusch2000sparse} for solving partial differential equations (PDEs). In the discretized matrices, the low-rank off-diagonal blocks arise because the associated Green's functions are smooth. The low-rank structured matrices also arise in applications that involve Toeplitz matrices (e.g., quantum chemistry, time-series analysis, queuing theory\ldots), etc. Identifying and compressing these low-rank blocks is the key to reducing the storage and computational costs of many matrix operations, such as solving linear systems, performing matrix-vector products, and computing eigenvalues.

Different algebraic \emph{low-rank representations} have been proposed in the literature. In particular, \(\mathcal{H}\)-matrices, \(\mathcal{H}^2\)-matrices, and Hierarchically Semi-Separable (HSS) matrices have been widely studied. It is not our goal to review these techniques and we recommend the references listed in~\cite{xia2010fast} for an overview. Some of these low-rank representations have been successfully implemented in software packages, but we are not aware of many publicly-available parallel libraries. In previous works, two codes based on the multifrontal method for solving sparse linear systems embedded HSS algorithms: Hsolver, a distributed-memory geometric code for finite-difference discretizations on regular meshes~\cite{wang2014parallel}, and StruMF, a sequential algebraic code~\cite{napov2015algebraic}. The other software packages that use low-rank approximation techniques include: Hlib (for \(\mathcal{H}\)- and \(\mathcal{H}^2\)-matrices)~\cite{borm1999h}, and MUMPS (sparse direct solver with Block Low-Rank approximation techniques)~\cite{amestoy2006hybrid,amestoy2014improving}.

\subsection{Contributions of this work}
Despite a large number of papers on the asymptotically low complexity of HSS-based representation and operations, the methods are mostly inaccessible to the high-performance computing community due to the lack of parallel software. Our work aims at providing a scalable package that can be used in large-scale applications. We have developed STRUMPACK - STRUctured Matrices PACKage - a package for computations with sparse and dense matrices. It combines HSS representations with a randomized sampling technique, which was not the case in our previous contributions. STRUMPACK has presently two main components: a distributed-memory dense matrix computations package and a shared-memory sparse direct solver. In this paper, we present the distributed-memory package. It is implemented using MPI and contains the following features:
\begin{itemize}
\item Compression into HSS form using randomized sampling.
\item Solving linear systems using ULV-like factorization and solution.
\item Computing HSS matrix-vector products.
\end{itemize}

STRUMPACK is a general package that does not make any assumption on the input matrix. It is algebraic (as opposed to geometric) and can work on any number of MPI processes. Our previously-developed geometric solver Hsolver also employs HSS compression and factorization kernels that can be used in a standalone way for dense matrices~\cite{wang2013efficient}. However, Hsolver is limited in usability -- it is a simplified code that works only with power-of-two number of processes and in single precision complex arithmetic. STRUMPACK does not have these limitations and, as presented here, it employs more recent algorithm advances (e.g., HSS combined with randomized sampling). It typically outperforms Hsolver, as we demonstrate in Section~\ref{sec:scal}.

In summary, the contributions of this work are the following:
\begin{itemize}
\item The library we present here (part of the STRUMPACK package) is the first randomized, distributed-memory, general purpose package for HSS matrix operations. It can use any number of MPI processes, not restricted to power-of-two as in Hsolver. It is up to 6x faster than the dense kernels used in Hsolver~\cite{wang2013efficient}.
\item We developed a flexible task-to-process mapping algorithm to accommodate non-uniform hierarchical matrix partitionings and unbalanced HSS trees. Therefore, the algorithms herein are fast for a wide range of applications (see Sections~\ref{sec:mapping} and~\ref{sec:comb}).
\item We developed an efficient parallel adaptive sampling method that is essential for problems with rank structures that cannot be estimated \emph{a priori}. This permits the solver to be used in a black-box fashion and increases its usability (see Section~\ref{sec:adaptive}).
\item We evaluated our algorithms for large-scale problems from a wide range of different academic and industrial applications, using large number of cores.
\end{itemize}

The rest of the paper is organized as follows. In Section~\ref{sec:HSS}, we review HSS techniques and the different ingredients of the HSS framework (HSS compression, ULV factorization and solution). In Section~\ref{sec:para}, we present our parallelization approach. We show how tasks are mapped and parallelized, we present our adaptive sampling mechanism, and we describe the communication features of our compression algorithm (number of messages and volume of communication). In Section~\ref{sec:exps}, we report on results using matrices from different applications. We show how HSS algorithms behave for different applications, and we present weak and strong scaling experiment to assess the performance of our code.

\section{Background on Hierarchically Semi-Separable matrices}
\label{sec:HSS}

We briefly introduce Hierarchically Semi-Separable (HSS) matrices. We mostly follow the notation used by Martinsson~\cite{martinsson2011fast}. We recommend~\cite{xia2010fast} for more theoretical aspects, and~\cite{xia2012superfast} for the use of HSS techniques for solving Toeplitz problems. The following references are works related to solving sparse linear systems using HSS techniques: \cite{xia2009superfast} (geometric setting, serial code), \cite{xia2013efficient} (algebraic setting, serial code), \cite{xia2013randomized} (algebraic setting, HSS techniques combined with randomized sampling, serial code), and \cite{wang2014parallel} (geometric setting, distributed-memory code).

\subsection{Representation}
\label{sec:HSSdef}

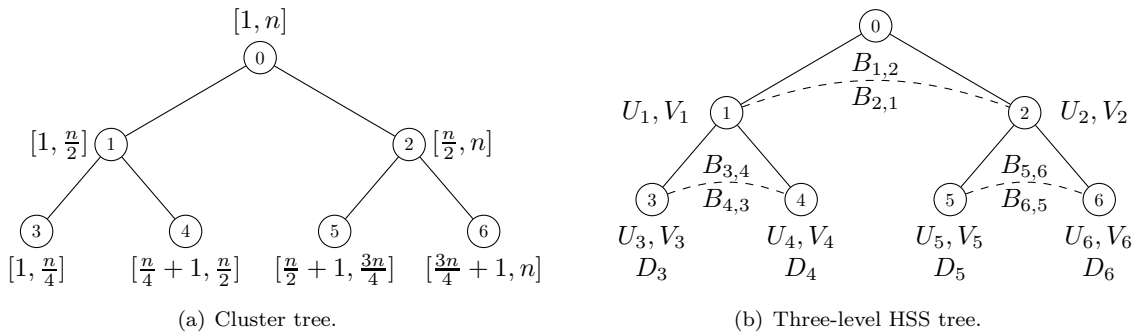
\begin{figure}[!ht]
\subfigure[Cluster tree.]{
\tikzstyle{cnode} = [draw, circle,scale=0.7]
\tikzstyle{level 1} = [level distance=.07\textwidth, sibling distance=.24\textwidth]
\tikzstyle{level 2} = [sibling distance=.12\textwidth]
\begin{tikzpicture}
  \node[cnode](0){0}
  child{node[cnode](1){1} child{node[cnode](3){3}} child{node[cnode](4){4}} }
  child{node[cnode](2){2} child{node[cnode](5){5}} child{node[cnode](6){6}} }  ;
  \node[above=.5em] at (0) {\([1             ,n]           \)};
  \node[left=.5em]  at (1) {\([1             ,\frac{n}{2}] \)};
  \node[right=.5em] at (2) {\([\frac{n}{2}   ,n]           \)};
  \node[below=.6em] at (3) {\([1             ,\frac{n}{4}] \)};
  \node[below=.6em] at (4) {\([\frac{n}{4}+1 ,\frac{n}{2}] \)};
  \node[below=.5em] at (5) {\([\frac{n}{2}+1 ,\frac{3n}{4}]\)};
  \node[below=.5em] at (6) {\([\frac{3n}{4}+1,n]           \)};
\end{tikzpicture}\\
}
\subfigure[Three-level HSS tree.]{
\tikzstyle{cnode} = [draw, circle,scale=0.7]
\tikzstyle{level 1} = [level distance=.07\textwidth, sibling distance=.24\textwidth]
\tikzstyle{level 2} = [sibling distance=.12\textwidth]
\begin{tikzpicture}
  \node[cnode](0){0}
  child{node[cnode](1){1} child{node[cnode](3){3}} child{node[cnode](4){4}} }
  child{node[cnode](2){2} child{node[cnode](5){5}} child{node[cnode](6){6}} }  ;
  \foreach \ln / \rn in {1/2,3/4,5/6}{
    \draw[dashed] (\ln) to [out=20,in=160] node[above=-0.3em]{\(B_{\ln,\rn}\)} node[below=-0.1em]{\(B_{\rn,\ln}\)} (\rn);
  }
  \foreach \ln in {3,4,5,6}{ \node[below=.5em] at (\ln) {\(\begin{array}{c}U_{\ln},V_{\ln}\\D_{\ln}\end{array}\)}; }
  \foreach \ln in {1}{  \node[left=.5em] at (\ln) {\(\begin{array}{c}U_{\ln},V_{\ln}\end{array}\)}; }
  \foreach \ln in {2}{ \node[right=.5em] at (\ln) {\(\begin{array}{c}U_{\ln},V_{\ln}\end{array}\)}; }
\end{tikzpicture}\\
}
\caption{Cluster tree and HSS tree associated with the example in Section~\ref{sec:HSSdef}.}
\label{fig:HSStree}
\end{figure}

HSS representations rely on a \emph{cluster tree} that defines a hierarchical clustering (or partitioning) of the index set \([1,n]\), where \(n\) is the number of rows and columns of the matrix we consider. A cluster tree is such that every node \(\tau\) is associated with an interval \(I_\tau\). The root node is associated with the interval \([1,n]\), and, for every node \(\tau\) of the tree with children \(\nu_1\) and \(\nu_2\), we have \(I_\tau=I_{\nu_1}\cup I_{\nu_2}\), and \(I_{\nu_1}\cap I_{\nu_2}=\emptyset \) (for simplicity we only consider binary trees, but the generalization is straightforward). The numbering of the nodes is done top-down; the root node is 0, and a node numbered \(i\) has children numbered \(2i+1\) and \(2i+2\). In Figure~\ref{fig:HSStree}(a), we show a possible cluster tree of \([1,n]\). In this example, the children of 2 are 5 and 6.

Any \(n\times n\) matrix \(A\) can be written into HSS form as follows:
\begin{enumerate}
\item Considering a \(2\times2\) partitioning of \(A\), i.e., a two-level cluster tree (one root node and two leaves), the off-diagonal blocks of \(A\) are decomposed into an ``SVD-like'' \(U B V\) form:
\begin{equation}
A=\begin{bmatrix}A_{1,1}  & A_{1,2} \\ A_{2,1} & A_{2,2}\end{bmatrix}
  = \begin{bmatrix}D_1   & U_1^{\text{big}} B_{1,2} {V_2^{\text{big}}}^* \\
U_2^{\text{big}} B_{2,1} {V_1^{\text{big}}}^* & D_2\end{bmatrix}
\label{eqn:1level_hss}
\end{equation}
The \(D\) matrices are simply the diagonal blocks of \(A\) and the \(U,B,V\) matrices are called \emph{generators}. We explain the reason of the ``\(^{\text{big}}\)'' superscript in the third point. \emph{This decomposition holds for any matrix \(A\), but it is useful in practice} (i.e., it reduces storage requirements and can be used for fast operations with \(A\)), \emph{only if the off-diagonal blocks of \(A\) are low-rank}. As mentioned in the introduction, this happens in many applications such as boundary element and finite element methods. When the off-diagonal blocks are low-rank, the \(U\) matrices are ``tall and skinny'', the \(B\) matrices are small and square or nearly square, and the \(V^*\) matrices are ``short and wide'', the aspect ratio depending on the ranks. The \(U,B,V\) matrices are computed using a rank-revealing factorization; we elaborate on this in the next sections.

Note that in situations where the off-diagonal blocks have large ranks, we may wish to \emph{approximate} them instead of computing their exact \(U B V\) decomposition; we show how in Section~\ref{sec:compr}. In this case, Equation (\ref{eqn:1level_hss}) provides us with an approximation of \(A\) that can be used, e.g., for preconditioning.

Note that this partitioning of the matrix corresponds to partitioning
\([1,n]\) as \([1,n]=I_1\cup I_2\).

\item Recursively, i.e., considering a three-level cluster tree, the off-diagonal blocks of the diagonal blocks of \(A\) are also decomposed into \(U,B,V\) form, and so on. After another stage of recursion:
\begin{equation}
A = \begin{bmatrix}\begin{bmatrix}D_3 & U_3^{\text{big}} B_{3,4} {V_4^{\text{big}}}^* \\
  U_4^{\text{big}} B_{4,3} {V_3^{\text{big}}}^* & D_4\end{bmatrix} & U_1^{\text{big}} B_{1,2} {V_2^{\text{big}}}^* \\
 U_2^{\text{big}} B_{2,1} {V_1^{\text{big}}}^* & \begin{bmatrix}D_5 & U_5^{\text{big}} B_{5,6} {V_6^{\text{big}}}^* \\
 U_6^{\text{big}} B_{6,5} {V_5^{\text{big}}}^* & D_6\end{bmatrix}\end{bmatrix}
\label{eqn:2level_hss}
\end{equation}
This partitioning corresponds to \(I_1=I_3\cup I_4\) and \(I_2=I_5\cup I_6\). \item There is a recursive relation between the generators appearing at different stages of recursions (which is the specificity of HSS and \(\mathcal{H}^2\)-matrices over the other classes of \(\mathcal{H}\)-matrices, and explains the use of the ``\(^{\text{big}}\)'' superscript):
\begin{equation}
U_1^{\text{big}}=\begin{bmatrix}U_3^{\text{big}} & 0 \\
 0 & U_4^{\text{big}}\end{bmatrix} U_1\ , \quad  V_1^{\text{big}}=\begin{bmatrix}V_3^{\text{big}} & 0 \\
 0 & V_4^{\text{big}}\end{bmatrix} V_1
\label{eqn:nested_basis}
\end{equation}
Thus,
\begin{equation}
A= \begin{bmatrix}\begin{bmatrix}D_3 & U_3^{\text{big}} B_{3,4} {V_4^{\text{big}}}^* \\
 U_4^{\text{big}} B_{4,3} {V_3^{\text{big}}}^* & D_4\end{bmatrix} & \; & \begin{bmatrix}U_3^{\text{big}} & 0 \\
 0 & U_4^{\text{big}}\end{bmatrix}U_1 B_{1,2} V_2^* \begin{bmatrix}{V_5^{\text{big}}}^* & 0 \\ 
 0 & {V_6^{\text{big}}}^*\end{bmatrix} \\ \\
\begin{bmatrix}U_5^{\text{big}} & 0 \\ 
  0 & U_6^{\text{big}}\end{bmatrix}U_2 B_{2,1} V_1^* \begin{bmatrix}{V_3^{\text{big}}}^* & 0 \\ 
  0 & {V_4^{\text{big}}}^*\end{bmatrix} & \; & \begin{bmatrix}D_5 & U_5^{\text{big}} B_{5,6} {V_6^{\text{big}}}^* \\
  U_6^{\text{big}} B_{6,5} {V_5^{\text{big}}}^* & D_6\end{bmatrix}\end{bmatrix}
\label{eqn:2level_hss_nested}
\end{equation}
This property is called the \emph{nested basis} property.
\end{enumerate}

In general, the HSS representation of \(A\) follows the structure of the cluster tree:
\begin{itemize}
\item For each leaf node \(\tau\), the corresponding diagonal block \(D_\tau=A(I_\tau,I_\tau)\) is left untouched (uncompressed, or ``full-rank'').
\item For each non-leaf node \(\tau\) with children \(\nu_1\) and \(\nu_2\), the corresponding off-diagonal blocks \(A_{\nu_1,\nu_2}=A(I_{\nu_1},I_{\nu_2})\) and \(A_{\nu_2,\nu_1}=A(I_{\nu_2},I_{\nu_1})\) are represented (exactly or approximately) by:%
\footnote{In the subsequent sections, when the context is clear, we will use equal sign
 instead of approximately equal.}
\begin{equation}
A_{\nu_1,\nu_2} \approx U_{\nu_1}^{\text{big}}B_{\nu_1,\nu_2}{V_{\nu_2}^{\text{big}}}^*
\label{eq:HSScompr}
\end{equation}
Furthermore, the hierarchical relation holds, i.e., basis are \emph{nested}:
\begin{equation}
U_\tau^{\text{big}}=\begin{bmatrix}U_{\nu_1}^{\text{big}} & 0 \\ 
0 & U_{\nu_2}^{\text{big}}\end{bmatrix} U_\tau \ , \quad
 V_\tau^{\text{big}}=\begin{bmatrix}V_{\nu_1}^{\text{big}} & 0 \\ 
0 & V_{\nu_2}^{\text{big}}\end{bmatrix} V_\tau
\end{equation}
\end{itemize}

Note that we never have to store or form explicitly the ``\(^{\text{big}}\)'' matrices at non-leaf nodes. Indeed, \(U\) at node \(\tau\) is given by \(U_\tau\) and the \(U_{\text{big}}\) matrices at its children \(\nu_1\) and \(\nu_2\), which are themselves given by looking at the grand-children of \(\tau\), and so on. At leaf nodes, \(U^{\text{big}}=U\). In Figure~\ref{fig:HSStree}(b), we show the tree corresponding to the previous example.

It is important to mention that the order of the rows and columns of matrix \(A\) matters. If \(A\) is shuffled randomly, the low-rank property is lost. In practice, matrices from real-life applications are often generated following an order that preserves the low-rank property. This was the case with all the matrices that we use in Section~\ref{sec:exps}. This point is developed in the literature~\cite{napov2015algebraic,wang2014parallel,amestoy2014improving}. 

In the rest of this section, we show how to obtain the HSS form of a matrix using randomized sampling. Then, we describe the different operations that can be performed with an HSS representation: matrix-vector product, ULV factorization (a specialized LU factorization), and triangular solution.

\subsection{Compression with randomized sampling}
\label{sec:compr}

Compression, i.e., construction of the HSS form of a matrix, is the most important algorithm of the HSS framework. Once the matrix is compressed, fast operations, such as a specialized factorization or specialized matrix-vectors products can be performed. We provide algorithmic details in the following sections.

The HSS compression algorithm we use is based on randomized sampling, which is essentially done by multiplying the input matrix with a set of random vectors. It was introduced by Martinsson~\cite{martinsson2011fast} and was also used by Xia et al. in a sparse multifrontal solver~\cite{xia2013randomized} and for algorithms for Toeplitz matrices~\cite{xia2012superfast}. The main advantage of this approach is that it does not require explicit access to all the entries of \(A\); it only requires a matrix-vector product routine and access to selected elements of \(A\). Therefore, \(A\) does not need to be explicitly formed, which saves memory, and the algorithm can benefit from an application-specific matrix-vector product. Furthermore, using randomized sampling simplifies the embedding of HSS kernels within a sparse solver~\cite{xia2013randomized}. This is the other component of the STRUMPACK project and is described in~\cite{ghysels2014multifrontal}.

Using a classical \(\mathcal{O}(n^2)\) matrix-vector product, the complexity of the compression operation is \(\mathcal{O}(r n^2)\) with \(r\) the maximum rank found during the compression, that we refer to as the \emph{HSS rank} of \(A\). In many applications, \(r\) is much smaller than \(n\). For example, it can be a small constant (e.g., 2D Poisson problems), or grow slowly with \(n\) (e.g., \(\log n\) for 2D Helmholtz or \(n^{1/3}\) for 3D Helmholtz problems)~\cite{xia2013efficient}. If a fast (typically \(\mathcal{O}(n)\)) matrix-vector product is available, the complexity drops to \(\mathcal{O}(r^2 n)\). Most of the floating-point operations happen when computing the samples, i.e., in the matrix-vector product. In a parallel setting, this helps load balancing in situations where very different ranks appear in different branches of the HSS tree.

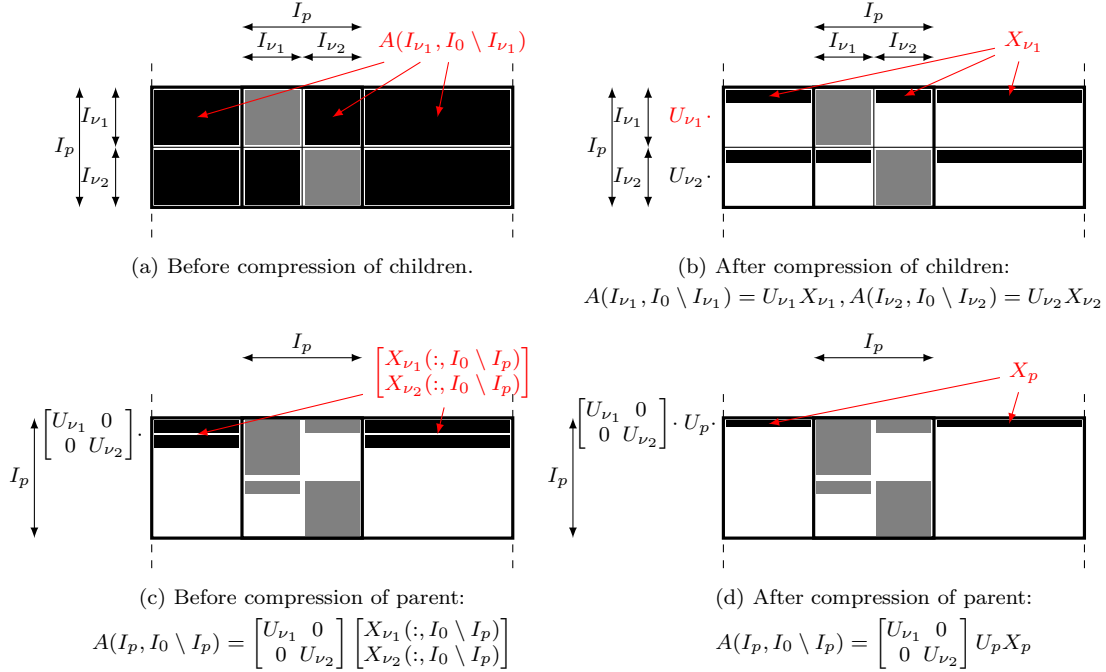
\begin{figure}[!ht]\centering\setlength{\arraycolsep}{0em}\tikzset{>=latex}\hspace{-0.7em}%
\begin{tikzpicture}[scale=0.4,font=\footnotesize]
\draw[<->] (-2.4,0)--(-2.4,4) node[midway,left=-0.2em]{\(I_p\)};
\draw[<->] (-1.2,0)--(-1.2,2) node[midway,left=-0.2em]{\(I_{\nu_2}\)};
\draw[<->] (-1.2,2)--(-1.2,4) node[midway,left=-0.2em]{\(I_{\nu_1}\)};
\draw[<->] (3,6)--(7,6) node[midway,above=-0.2em]{\(I_p\)};
\draw[<->] (3,5)--(5,5) node[midway,above=-0.2em]{\(I_{\nu_1}\)};
\draw[<->] (5,5)--(7,5) node[midway,above=-0.2em]{\(I_{\nu_2}\)};
\fill[gray](3.1,2.1)--(4.9,2.1)--(4.9,3.9)--(3.1,3.9)--cycle;
\fill[gray](5.1,0.1)--(6.9,0.1)--(6.9,1.9)--(5.1,1.9)--cycle;
\fill(0.1,0.1)--(2.9,0.1)--(2.9,1.9)--(0.1,1.9)--cycle;
\fill(3.1,0.1)--(4.9,0.1)--(4.9,1.9)--(3.1,1.9)--cycle;
\fill(7.1,0.1)--(11.9,0.1)--(11.9,1.9)--(7.1,1.9)--cycle;
\fill(5.1,2.1)--(6.9,2.1)--(6.9,3.9)--(5.1,3.9)--cycle;
\fill(7.1,2.1)--(11.9,2.1)--(11.9,3.9)--(7.1,3.9)--cycle;
\fill(0.1,2.1)--(2.9,2.1)--(2.9,3.9)--(0.1,3.9)--cycle;
\draw[dashed](0,-1)--(0,0);
\draw[dashed](12,-1)--(12,0);
\draw[dashed](12,4)--(12,5);
\draw[dashed](0,4)--(0,5);
\draw[very thick](0,0)--(12,0)--(12,4)--(0,4)--cycle;
\draw[very thick](3,0)--(7,0)--(7,4)--(3,4)--cycle;
\draw(0,2)--(12,2);
\draw(5,0)--(5,4);
\node at (5,-2){(a) Before compression of children.};
\node[red](1) at (10,5.5) {\(A(I_{\nu_1},I_0\setminus I_{\nu_1})\)};
\draw[red,->](1) to (1.5,3);
\draw[red,->](1) to (6,3);
\draw[red,->](1) to (9.5,3);
\begin{scope}[shift={(19,0)}]
\draw[<->] (-3.7,0)--(-3.7,4) node[midway,left=-0.2em]{\(I_p\)};
\draw[<->] (-2.5,0)--(-2.5,2) node[midway,left=-0.2em]{\(I_{\nu_2}\)};
\draw[<->] (-2.5,2)--(-2.5,4) node[midway,left=-0.2em]{\(I_{\nu_1}\)};
\draw[<->] (3,6)--(7,6) node[midway,above=-0.2em]{\(I_p\)};
\draw[<->] (3,5)--(5,5) node[midway,above=-0.2em]{\(I_{\nu_1}\)};
\draw[<->] (5,5)--(7,5) node[midway,above=-0.2em]{\(I_{\nu_2}\)};
\node at (-1.1,1){\(U_{\nu_2}\cdot\)};
\node[red] at (-1.1,3){\(U_{\nu_1}\cdot\)};
\fill[gray](3.1,2.1)--(4.9,2.1)--(4.9,3.9)--(3.1,3.9)--cycle;
\fill[gray](5.1,0.1)--(6.9,0.1)--(6.9,1.9)--(5.1,1.9)--cycle;
\fill(0.1,1.5)--(2.9,1.5)--(2.9,1.9)--(0.1,1.9)--cycle;
\fill(3.1,1.5)--(4.9,1.5)--(4.9,1.9)--(3.1,1.9)--cycle;
\fill(7.1,1.5)--(11.9,1.5)--(11.9,1.9)--(7.1,1.9)--cycle;
\fill(5.1,3.5)--(6.9,3.5)--(6.9,3.9)--(5.1,3.9)--cycle;
\fill(7.1,3.5)--(11.9,3.5)--(11.9,3.9)--(7.1,3.9)--cycle;
\fill(0.1,3.5)--(2.9,3.5)--(2.9,3.9)--(0.1,3.9)--cycle;
\draw[dashed](0,-1)--(0,0);
\draw[dashed](12,-1)--(12,0);
\draw[dashed](12,4)--(12,5);
\draw[dashed](0,4)--(0,5);
\draw[very thick](0,0)--(12,0)--(12,4)--(0,4)--cycle;
\draw[very thick](3,0)--(7,0)--(7,4)--(3,4)--cycle;
\draw(0,2)--(12,2);
\draw(5,0)--(5,4);
\node at (4,-2){(b) After compression of children:};
\node at (4,-3){\(A(I_{\nu_1},I_0\setminus I_{\nu_1})=U_{\nu_1}X_{\nu_1},A(I_{\nu_2},I_0\setminus I_{\nu_2})=U_{\nu_2}X_{\nu_2}\)};
\node[red](1) at (10,5.5) {\(X_{\nu_1}\)};
\draw[red,->](1) to (1.5,3.7);
\draw[red,->](1) to (6,3.7);
\draw[red,->](1) to (9.5,3.7);
\end{scope}
\begin{scope}[shift={(0,-11)}]
\draw[<->] (-3.9,0)--(-3.9,4) node[midway,left=-0.2em]{\(I_p\)};
\draw[<->] (3,6)--(7,6) node[midway,above=-0.2em]{\(I_p\)};
\node at (-2.0,3.45){\(\begin{bmatrix} U_{\nu_1} & 0\\ 0 & U_{\nu_2}\end{bmatrix}\!\cdot\)};
\fill[gray](3.1,2.1)--(4.9,2.1)--(4.9,3.9)--(3.1,3.9)--cycle;
\fill[gray](5.1,0.1)--(6.9,0.1)--(6.9,1.9)--(5.1,1.9)--cycle;
\fill(0.1,3.0)--(2.9,3.0)--(2.9,3.4)--(0.1,3.4)--cycle;
\fill[gray](3.1,1.5)--(4.9,1.5)--(4.9,1.9)--(3.1,1.9)--cycle;
\fill(7.1,3.0)--(11.9,3.0)--(11.9,3.4)--(7.1,3.4)--cycle;
\fill[gray](5.1,3.5)--(6.9,3.5)--(6.9,3.9)--(5.1,3.9)--cycle;
\fill(7.1,3.5)--(11.9,3.5)--(11.9,3.9)--(7.1,3.9)--cycle;
\fill(0.1,3.5)--(2.9,3.5)--(2.9,3.9)--(0.1,3.9)--cycle;
\draw[dashed](0,-1)--(0,0);
\draw[dashed](12,-1)--(12,0);
\draw[dashed](12,4)--(12,5);
\draw[dashed](0,4)--(0,5);
\draw[very thick](0,0)--(12,0)--(12,4)--(0,4)--cycle;
\draw[very thick](3,0)--(7,0)--(7,4)--(3,4)--cycle;
\node at (5,-2){(c) Before compression of parent:};
\node at (5,-3.5){\(A(I_p,I_0\setminus I_p)= \begin{bmatrix} U_{\nu_1} & 0\\ 0 & U_{\nu_2}\end{bmatrix} \begin{bmatrix}X_{\nu_1}(:,I_0\setminus I_p)\\X_{\nu_2}(:,I_0\setminus I_p)\end{bmatrix}\)};
\node[red](1) at (10,5.5) {\(\begin{bmatrix}X_{\nu_1}(:,I_0\setminus I_p)\\X_{\nu_2}(:,I_0\setminus I_p)\end{bmatrix}\)};
\draw[red,->](1) to (1.5,3.45);
\draw[red,->](1) to (9.5,3.45);
\end{scope}
\begin{scope}[shift={(19,-11)}]
\draw[<->] (-5.0,0)--(-5.0,4) node[midway,left=-0.2em]{\(I_p\)};
\draw[<->] (3,6)--(7,6) node[midway,above=-0.2em]{\(I_p\)};
\node at (-2.6,3.8){\(\begin{bmatrix} U_{\nu_1} & 0\\ 0 & U_{\nu_2}\end{bmatrix}\!\cdot U_p\cdot\)};
\fill[gray](3.1,2.1)--(4.9,2.1)--(4.9,3.9)--(3.1,3.9)--cycle;
\fill[gray](5.1,0.1)--(6.9,0.1)--(6.9,1.9)--(5.1,1.9)--cycle;
\fill[gray](3.1,1.5)--(4.9,1.5)--(4.9,1.9)--(3.1,1.9)--cycle;
\fill[gray](5.1,3.5)--(6.9,3.5)--(6.9,3.9)--(5.1,3.9)--cycle;
\fill(7.1,3.7)--(11.9,3.7)--(11.9,3.9)--(7.1,3.9)--cycle;
\fill(0.1,3.7)--(2.9,3.7)--(2.9,3.9)--(0.1,3.9)--cycle;
\draw[dashed](0,-1)--(0,0);
\draw[dashed](12,-1)--(12,0);
\draw[dashed](12,4)--(12,5);
\draw[dashed](0,4)--(0,5);
\draw[very thick](0,0)--(12,0)--(12,4)--(0,4)--cycle;
\draw[very thick](3,0)--(7,0)--(7,4)--(3,4)--cycle;
\node at (5,-2){(d) After compression of parent:};
\node at (5,-3.5){\(A(I_p,I_0\setminus I_p)= \begin{bmatrix} U_{\nu_1} & 0\\ 0 & U_{\nu_2}\end{bmatrix} U_p X_p\)};
\node[red](1) at (10,5.5) {\(X_p\)};
\draw[red,->](1) to (1.5,3.8);
\draw[red,->](1) to (9.5,3.8);
\end{scope}
\end{tikzpicture}
\caption{Compression process without randomized sampling, at two child nodes \(\nu_1\) and \(\nu_2\) and their parent \(\tau\). Full blocks are off-diagonal blocks to be compressed. Shaded blocks are that are left untouched.}
\label{fig:HSScompr}
\end{figure}

We briefly recall how HSS compression \emph{without} randomized sampling works, as described in~\cite{xia2010fast}. The main property that we use is that, at each node \(\tau\), the off-diagonal row blocks and column blocks \(A(I_\tau,I_0\setminus I_\tau)\) and \(A(I_0\setminus I_\tau,I_\tau)\) are low-rank, denoting \(I_0=[1,n]\). These blocks are referred to as the \emph{strip row Hankel blocks} and \emph{strip column Hankel blocks} of \(A\) in~\cite{chandrasekaran2010numerical}. Consider row blocks. We traverse the tree following a postorder, from the leaf nodes up to the root node. At a leaf node \(l\), \(A(I_l,I_0\setminus I_l)\) is low rank and we can find a basis \(U_l\) for the rows, by using a rank revealing factorization: \(A(I_l,I_0\setminus I_l)=U_l X_l\). At the parent node \(p\), we wish to compress \(A(I_p,I_0\setminus I_p)\). However compressing this block directly is potentially
expensive and does not make use of the nested basis property.
 Instead, we use:
\begin{equation}
A(I_p,I_0\setminus I_p)  = \begin{bmatrix} A(I_{\nu_1},I_0\setminus I_p)\\ A(I_{\nu_2},I_0\setminus I_p)\end{bmatrix}
                         = \begin{bmatrix} U_{\nu_1} X_{\nu_1}(:,I_0\setminus I_p) \\U_{\nu_2} X_{\nu_2}(:,I_0\setminus I_p)\end{bmatrix}\\
                         = \begin{bmatrix} U_{\nu_1} & 0\\ 0 & U_{\nu_2}\end{bmatrix} \begin{bmatrix}X_{\nu_1}(:,I_0\setminus I_p)\\X_{\nu_2}(:,I_0\setminus I_p)\end{bmatrix}
\end{equation}
Our objective is to compress \(A(I_p,I_0\setminus I_p)\) as \(A(I_p,I_0\setminus I_p)=U_p^{\text{big}}X_p\); using the above equation, we get \(U_p^{\text{big}}\) and \(X_p\) by computing a rank revealing factorization of \(\begin{bmatrix}X_{\nu_1}(:,I_0\setminus I_p)\\X_{\nu_2}(:,I_0\setminus I_p)\end{bmatrix}\), instead of compressing \(A(I_p,I_0\setminus I_p)\) directly. This process is illustrated in Figure~\ref{fig:HSScompr}. Column blocks are compressed in a similar way to obtain the \(V\) generators, using the \(X\) obtained during the compression of row blocks.

The randomized compression algorithm follows a similar process, except that it relies on samples of the input matrix instead of accessing the matrix directly. For now we suppose that the maximum rank \(r\) is known a priori. We relax this assumption in Section~\ref{sec:adaptive}. Let \(R^r\) and \(R^c\) be \(n \times d\) tall and skinny random matrices with \(d = r + p\) columns, where \(p\) is a small oversampling parameter (Martinsson recommends \(p=10\)). Let \(S^r=AR^r\) and \(S^c=A^*R^c\) be samples for the row and column bases of \(A\) respectively. For a non-leaf node \(\tau\) with children \(\nu_1\) and \(\nu_2\), let \(D_\tau\) be defined as
\[D_\tau = \begin{bmatrix} D_{\nu_1} & A_{\nu_1,\nu_2} \\ A_{\nu_2,\nu_1} & D_{\nu_2} \end{bmatrix}\]
If \(\{\tau_1,\tau_2,\dots,\tau_q\}\) are the nodes at level \(\ell\) of the HSS tree, then
\[D^{(\ell)} = \text{diag}(D_{\tau_1}, D_{\tau_2},\dots,D_{\tau_q})\]
is an \(n \times n\) block diagonal matrix. The main idea of the randomized sampling algorithm is to construct a row sample matrix \(S^{(\ell)}\) for each level of the tree as
\[S^{(\ell)} = \left( A - D^{(\ell)} \right) R^r = S^r - D^{(\ell)} R^r\]
This row sample matrix \(S^{(\ell)}\) captures the action of a product of the block off-diagonal part of \(A\) with a set of random vectors \(R^r\). It is exactly this block off-diagonal part that needs to be compressed using low-rank approximation to obtain the HSS generators. Similarly, we compute a sample matrix using \(S^c\) and \(R^c\) to capture the column space of the off-diagonal blocks.

A central component of the randomized sampling algorithm is the Interpolative Decomposition (ID)~\cite{cheng2005compression}. The ID computes a factorization of a rank-\(k\) \(m\times n\) matrix \(Y\) by expressing the columns of \(Y\) as linear combinations of a subset of columns of \(Y\):
\[\left[X,J\right]=\text{ID}(Y), \text{ s.t. } Y=Y(:,J)X \text{ where } Y \text{ is } m\times k \text{ and } X \text{ is } k\times n\]
A compression tolerance \(\varepsilon\) can be added as a parameter:
\[\left[X,J\right]=\text{ID}(Y,\varepsilon), \text{ s.t. } Y\simeq Y(:,J)X \text{ where } Y \text{ is } m\times k' \text{ and } X \text{ is } k'\times n\]
where the numerical rank \(k' \leq k\). The ID can be computed using, for example, a \(QR\) factorization with column pivoting~\cite{chan1987rank,quintana1998blas}
\begin{align*}
Y &= Q\,R\,\Pi^{-1}\hspace{8em}(\Pi\text{: permutation matrix representing column pivoting})\\
  &= Q\,\left[R_1\ \ R_2\right]\Pi^{-1}\hspace{5em}(R_1: k \times k)\\
  &= (Q\,R_1)\,(\left[I\ \ R_1^{-1}R_2\right]\Pi^{-1})\\
  &= Y(:,J)\,X\hspace{7.8em}(Q\,R_1\text{: first columns of pivoted }Y)
\end{align*}

A consequence of using Interpolative Decomposition is that \(B_{\nu_1,\nu_2} = A(I^r_{\nu_1}, I^c_{\nu_2})\) is a submatrix of the original matrix \(A\). Furthermore, it also leads to a special structure for the \(U_\tau\) and \(V_\tau\) generators:
\[U_\tau = \Pi^r_\tau \begin{bmatrix} I \\ E^r_\tau \end{bmatrix}\quad
 \text{ and }\quad V_\tau = \Pi^c_\tau \begin{bmatrix} I \\ E^c_\tau \end{bmatrix}\]
where \(U_\tau\) and \(V_\tau\) have respective column ranks \(r_\tau^r\) and \(r_\tau^c\), \(\Pi_\tau^r\) and \(\Pi_\tau^c\) are permutation matrices and the \(I\)'s are the Identity matrices, one is of order \(r_\tau^r\), the other of order \(r_\tau^c\). This structure is exploited in the factorization, as shown in Section~\ref{sec:ULV}, and it allows for faster operations with the generators. From a memory viewpoint, we only need to store the \(E\) matrices, and the permutation matrices \(\Pi\) are represented by a single vector. A remarkable consequence is that when the block we want to compress is full-rank, the generators have the degenerate form \(U_\tau=\Pi^r_\tau I\), and therefore they can be stored at very low cost (only the permutation information needs to be stored).

The compression algorithm works as follows:
\begin{enumerate}
\item Generate \(R^r\) and \(R^c\) random \(n\times d\) matrices.
\item Compute the samples \(S^r=AR^r\) and \(S^c=A^*R^c\).
\item Traverse the tree in topological order (i.e., children before parents): at each node,
\begin{enumerate}
\item Construct local samples.
\item Compute generators using Interpolative Decomposition.
\item Update samples and random vectors to make the construction of local samples faster at subsequent nodes.
\end{enumerate}
\end{enumerate}
The detailed algorithm is presented in Algorithm~\ref{alg:HSScompr}. Note that in the serial case, the topological order that we follow is simply a postordering of the HSS tree. However, in the parallel case, we follow a more general topological order, as described in Section~\ref{sec:mapping}. Note that the Interpolative Decomposition (step (3)(b), line~\ref{algline:ID} in the algorithm) is the step where the user-given threshold \(\varepsilon\) is used. The \(QR\) factorization with column pivoting stops when \(\frac{R_{ii}}{R_{11}}\leq\varepsilon\).

\begin{algorithm}[!ht]
\KwData{\(d = r + 10\) with \(r\) an upper bound for the rank of \(A \in \mathbb{R}^{n \times n}\)\newline
  \(S^r = AR^r\) and \(S^c = A^* R^c\) with \(\{ S^r, S^c, R^r, R^c \} \in \mathbb{R}^{n \times d}\)\newline
  A tree on the index vector \([1,n]\) with an index set \(I_\tau\) at each node \(\tau\)
}
\KwResult{Basis matrices defining the HSS matrix:\newline
\(D_\tau\) at the leaves, \(U_\tau\), \(V_\tau\) at all nodes except the root\newline
\(B_{\nu_1, \nu_2}\) at non-leaves for all children combinations}
\BlankLine
\ForEach{node \(\tau\) in topological order (bottom-up traversal)}{
  \eIf{node \(\tau\) is a leaf}{
    \(D_\tau = A(I_\tau, I_\tau)\) \\
    \mytab{\(S^r_{\text{loc}} = S^r(I_\tau,:) - D_\tau   R^r(I_\tau,:)\)}
          {\(S^c_{\text{loc}} = S^c(I_\tau,:) - D_\tau^* R^c(I_\tau,:)\)}
  }{
    Let \(\nu_1\) and \(\nu_2\) be the two children of node \(\tau\)\\
    \mytab{\(B_{\nu_1,\nu_2} = A(I^r_{\nu_1}, I^c_{\nu_2})\)}
          {\(B_{\nu_2,\nu_1} = A(I^r_{\nu_2}, I^c_{\nu_1})\)}\\
    \mytab{\(S^r_{\text{loc}} = \begin{bmatrix} S^r_{\nu_1} - B_{\nu_1,\nu_2} R^r_{\nu_2}\\
                                                  S^r_{\nu_2} - B_{\nu_2,\nu_1} R^r_{\nu_1}
                                  \end{bmatrix}\)}
          {\(S^c_{\text{loc}} = \begin{bmatrix} S^c_{\nu_1} - B^*_{\nu_2,\nu_1} R^c_{\nu_2}\\
                                                  S^c_{\nu_2} - B^*_{\nu_1,\nu_2} R^c_{\nu_1}
                                  \end{bmatrix}\)}
  }
  \mytab{\(\begin{bmatrix} (U_\tau)^*, J^r_\tau \end{bmatrix} = \textbf{ID}\left((S^r_{\text{loc}})^* \right)\)}
        {\(\begin{bmatrix} (V_\tau)^*, J^c_\tau \end{bmatrix} = \textbf{ID}\left((S^c_{\text{loc}})^* \right)\)}~\label{algline:ID}\\
  \mytab{\(S^r_\tau = S^r_{\text{loc}}(J^r_\tau,:)\)}
        {\(S^c_\tau = S^c_{\text{loc}}(J^c_\tau,:)\)}\\
  \eIf{node \(\tau\) is a leaf}{
    \mytab{\(R^r_\tau = (V_\tau)^* R^r(I_\tau,:)\)}
          {\(R^c_\tau = (U_\tau)^* R^c(I_\tau,:)\)}\\
    \mytab{\(I^r_\tau = I_{\tau}(J^r_\tau)\)}
          {\(I^c_\tau = I_{\tau}(J^c_\tau)\)}\\
  }{
    \mytab{\(R^r_\tau = (V_\tau)^* \begin{bmatrix} R^r_{\nu_1} \\ R^r_{\nu_2}\end{bmatrix}\)}
          {\(R^c_\tau = (U_\tau)^* \begin{bmatrix} R^c_{\nu_1} \\ R^c_{\nu_2}\end{bmatrix}\)}\\
    \mytab{\(I^r_\tau = [I^r_{\nu_1} \,\, I^r_{\nu_2} ](J^r_\tau)\)}
          {\(I^c_\tau = [I^c_{\nu_1} \,\, I^c_{\nu_2} ](J^c_\tau)\)}
  }
}
\caption{Computing the HSS representation of an unsymmetric matrix.}
\label{alg:HSScompr}
\end{algorithm}

\subsection{Matrix-vector product}\label{sec:matvec}
Once a matrix is compressed into an HSS form, matrix-vector products can be computed in \(\mathcal{O}(rn)\), thus typically faster than using a classical \(\mathcal{O}(n^2)\) product. However, the compression cost is \(\mathcal{O}(r n^2)\) using a standard non-randomized algorithm, or using a randomized algorithm based on samples computed with standard matrix-vector products; therefore, it is amortized only when multiple products are computed, either successively or with blocks of vectors. This is the case for example in iterative linear solvers or eigensolvers. The HSS matrix-vector algorithm consists of two traversals of the HSS tree, as shown in Algorithm~\ref{alg:prod}. The first traversal accumulates the actions of the \(V\) generators, while the other traversal uses the \(U\) generators as well as the \(B_{\nu_1,\nu_2}\), \(B_{\nu_2,\nu_1}\) and \(D_\tau\) matrices.

\begin{algorithm}[!ht]
\KwData{HSS form: \(D_\tau\) (leaves), \(U_\tau\), \(V_\tau\) (all nodes except root), \(B_{\nu_1, \nu_2}\) and \(B_{\nu_2, \nu_1}\) (non-leaves).\newline
  Right-hand side \(x\) (one or more columns).}
\KwResult{\(b=Ax\).}
\BlankLine
\ForEach{node \(\tau\) in topological order (bottom-up traversal)}{
  \eIf{node \(\tau\) is a leaf}{
    \(y_\tau = V_\tau^* x(I_\tau,:)\)
  }{
    \(y_\tau = V_\tau^*\begin{bmatrix}y_{\nu_1}\\ y_{\nu_2}\end{bmatrix}\)
  }
}
\(z_\tau=0\) for root node\\
\ForEach{node \(\tau\) in reverse topological order (top-down traversal)}{
  \eIf{node \(\tau\) is a leaf}{
    \(b(I_\tau,:)=U_\tau z_\tau+D_\tau x(I_\tau,:)\)
  }{
    \(\begin{bmatrix}z_{\nu_1}\\ z_{\nu_2}\end{bmatrix}=\begin{bmatrix}0 & B_{\nu_1,\nu_2}\\ B_{\nu_2,\nu_1} & 0\end{bmatrix}\begin{bmatrix}y_{\nu_1}\\ y_{\nu_2}\end{bmatrix}+U_\tau z_\tau\)
  }
}
\caption{HSS matrix-vector product, for a non-symmetric matrix.}
\label{alg:prod}
\end{algorithm}

\subsection{ULV-like factorization}
\label{sec:ULV}
A matrix in HSS form can be factored using a special form of factorization called ULV factorization~\cite{chandrasekaran2006fast}. Then, the factored form can be used to obtain the solution to the linear system. We now describe the factorization algorithm, using a two-stage HSS example (i.e., a three-level tree) to aid exposition.

In the original ULV factorization, fast orthogonal transformations are used to eliminate \(\mathcal{O}(n-r)\) unknowns; the remaining \(\mathcal{O}(r)\) unknowns are eliminated using a standard LU factorization. The factorization we use does not use orthogonal transformations but instead it exploits the special structure of the HSS generators that comes from the Interpolative Decomposition. Algorithm~\ref{alg:ULV} shows the complete ULV factorization procedure. In the following we explain how it works, starting from the one-stage HSS form~(\ref{eqn:1level_hss}), i.e., a two-level tree.

Recall that each \(U\) generator has the special structure 
\(U_\tau= \Pi^r_\tau \begin{bmatrix} I \\ E^r_\tau \end{bmatrix}\).
Define \(\Omega_\tau=\begin{bmatrix}-E^r_\tau & I \\ I & 0\end{bmatrix}{\Pi^r_\tau}^{T}\).
Then the transformation
 \(\Omega_\tau U_\tau=\begin{bmatrix}0 \\ I\end{bmatrix}\)
introduces a zero block on the top, where \(I\) is of order \(r_\tau^r\). Now consider the one-stage HSS decomposition (as in Equation (\ref{eqn:1level_hss})):
\[A=\begin{bmatrix}A_{1,1} & A_{1,2}\\A_{2,1} & A_{2,2}\end{bmatrix}=\begin{bmatrix}D_1 & U_1 B_{1,2} V_2^*\\U_2 B_{2,1} V_1^* & D_2\end{bmatrix}\]
Applying \(\Omega_1\) and \(\Omega_2\), we get:
\[\begin{bmatrix}\Omega_1 & 0\\ 0 & \Omega_2\end{bmatrix}A=\begin{bmatrix}\Omega_1 D_1 & \begin{bmatrix}0\\B_{1,2}V_2^*\end{bmatrix}\\ \begin{bmatrix}0\\B_{2,1}V_1^*\end{bmatrix}&\Omega_2 D_2\end{bmatrix}\]
At each node \(\tau\), we partition \(W_\tau=\Omega_\tau D_\tau\) into the top (t) and bottom (b) parts, \(W_\tau=\begin{bmatrix}W_{\tau;t}\\W_{\tau;b}\end{bmatrix}\) where \(W_{\tau;b}\) has \(r_\tau^r\) rows, and we perform an \(LQ\) decomposition of \(W_{\tau;t}\), \(W_{\tau;t}=\left[L_\tau\ 0\right]Q_\tau\). Then,
\begin{align}
\begin{bmatrix}\Omega_1 &  \\   & \Omega_2\end{bmatrix} A 
\begin{bmatrix}Q_1^* &  \\  & Q_2^*\end{bmatrix} 
   &=
 \begin{bmatrix}\begin{bmatrix}L_1 & 0\end{bmatrix}& 0\\W_{1;b}Q_1^*&B_{1,2}V_2^*Q_2^*\\0&\begin{bmatrix}L_2 & 0\end{bmatrix}\\B_{2,1}V_1^*Q_1^*&W_{2;b}Q_2^*\end{bmatrix} \nonumber \\
   &=
\begin{bmatrix}L_1 & 0 & 0 & 0   \\
  W_{1;b}Q_{1;t}^*  & \underline{W_{1;b}Q_{1;b}^*}
         & B_{1,2}V_2^*Q_{2;t}^*    & \underline{B_{1,2}V_2^*Q_{2;b}^*}  \\ 
  0 & 0 & L_2 & 0 \\
  B_{2,1}V_1^*Q_{1;t}^*   &  \underline{B_{2,1}V_1^*Q_{1;b}^*} 
         & W_{2;b}Q_{2;t}^*         & \underline{W_{2;b}Q_{2;b}^*}
\end{bmatrix}
\label{eqn:1level_ulv_transform}
\end{align}
\emph{Implicitly}, if we swap block rows (and columns) corresponding to the \{1;b\} and \{2;t\} parts, denoted by a permutation matrix \(\Gamma_{1;b\leftrightarrow 2;t} = \left[\begin{smallmatrix} I &&& \\ & 0 & I & \\ & I & 0 & \\ &&& I \end{smallmatrix}\right]\) the above transformation can be written in the \(U L V\) factored form:
\begin{equation}
A =
\underbrace{\begin{bmatrix}\Omega_1^{-1} & \\  & \Omega_2^{-1}\end{bmatrix}
\Gamma_{1;b\leftrightarrow 2;t}}_{\text{\large\(U\)}} \cdot
\underbrace{\begin{bmatrix} L_1  &  & \\ 0 & L_2 & \\ L_{2,1} & L_{1,2} & D_0 \end{bmatrix}}_{\text{\large\(L\)}} \cdot 
\underbrace{\Gamma_{1;b\leftrightarrow 2;t}^T
  \begin{bmatrix} Q_1  &  \\   & Q_2 \end{bmatrix} }_{\text{\large\(V\)}}
\label{eqn:1level_ulv}
\end{equation}
where \(L_{2,1}=\begin{bmatrix} W_{1;b}Q_{1;t}^* \\ B_{2,1}V_1^*Q_{1;t}^*\end{bmatrix}\) and
\(L_{1,2}=\begin{bmatrix} B_{1,2}V_2^*Q_{2;t}^* \\ W_{2;b}Q_{2;t}^* \end{bmatrix}\),
and \(D_0\) is the reduced submatrix
\[D_0 \stackrel{\mathrm{def}}{=}
\begin{bmatrix} W_{1;b}Q_{1;b}^*&B_{1,2}V_2^*Q_{2;b}^* \\ 
B_{2,1}V_1^*Q_{1;b}^*&W_{2;b}Q_{2;b}^* \end{bmatrix}
\stackrel{\mathrm{def}}{=}
\begin{bmatrix} \tilde{D}_1   & B_{1,2}V_2^*Q_{2;b}^* \\ 
  B_{2,1}V_1^*Q_{1;b}^*       & \tilde{D}_2 \end{bmatrix}
\]

This is how the name ``ULV-factorization'' came from~\cite{chandrasekaran2006fast}. In the original form, both ``U'' and ``V'' transformations are orthogonal. But here, since \(\Omega_\tau\) has the special structure stemming from the Interpolative Decomposition, it may not be orthogonal. Therefore, we refer to it as a ``ULV-like'' factorization. Note that factorization (\ref{eqn:1level_ulv}) is not used in the solution procedure; instead, it is the transformation (\ref{eqn:1level_ulv_transform}) that is actually used, as we show in the next section. This transformation is also used in the sparse factorization by Xia~\cite{xia2013randomized}.

With the ``L'' form above, the unknowns corresponding to \(L_1\) and \(L_2\) can be eliminated using a regular forward substitution. The reduced submatrix \(D_0\) corresponds to the \(\mathcal{O}(r)\) remaining unknowns, and is formed at the parent node, which is also a root node, where \(LU(D_0)\) is performed. This is illustrated in Figure~\ref{fig:ULV}, with \(D_0\) colored red.

\begin{figure}[!ht]
\centering
\begin{tikzpicture}[scale=0.04]
\newcommand{\myrows}[1][]{\fill[#1] (0,0) rectangle +(16,4);\fill[#1,red] (17,0) rectangle +(4,4);}
\newcommand{\mytriangle}[1][]{\fill[#1] (0,0) -- +(16,0) -- +(0,16);}
\newcommand{\myUBV}[1][]{\fill[#1] (0,0) rectangle +(4,16.5);\fill[#1] (0,17) rectangle +(4,4);\fill[#1] (4.5,17) rectangle +(16.5,4);}
\newcommand{\mylbracket}[1][]{\draw[thick,#1] (0,0) -- ++(-2,0) -- ++(0,47) -- ++(2,0);}
\newcommand{\myrbracket}[1][]{\draw[thick,#1] (0,0) -- ++(2,0) -- ++(0,47) -- ++(-2,0);}
\mylbracket[shift={(0,-2)}]
\myrbracket[shift={(43,-2)}]
\fill (22,0) rectangle +(21,21);
\fill (0,22) rectangle +(21,21);
\myUBV
\myUBV[shift={(22,22)}]
\node at (29.5,25) {\footnotesize\(U_1\)};
\node at (39,35) {\footnotesize\(V_2\)};
\node at (32.5,31) {\footnotesize\(B_{1,2}\)};
\draw[->] (32,34) -- (27,38.5);
\draw[thick,->](54,21.5) -> +(46,0);
\node at (75,10){\footnotesize \(\begin{bmatrix}\Omega_1 & 0\\ 0 & \Omega_2\end{bmatrix}\times\)\ldots};
\begin{scope}[shift={(110,0)}]
\mylbracket[shift={(0,-2)}]
\myrbracket[shift={(43,-2)}]
\fill (0,0) rectangle +(21,4);
\fill (22,0) rectangle +(21,21);
\fill (0,22) rectangle +(21,21);
\fill (22,22) rectangle +(21,4);
\draw[thick,->](54,21.5) -> +(46,0);
\node at (75,10){\footnotesize \(\ldots\times\begin{bmatrix}Q_1^* & 0\\ 0 & Q_2^*\end{bmatrix}\)};
\end{scope}
\begin{scope}[shift={(220,0)}]
\mylbracket[shift={(0,-2)}]
\myrbracket[shift={(43,-2)}]
\myrows
\myrows[shift={(22,0)}]
\mytriangle[shift={(22,5)}]
\myrows[shift={(0,22)}]
\myrows[shift={(22,22)}]
\mytriangle[shift={(0,27)}]
\draw[thick,->](54,21.5) -> +(46,0);
\node at (75,10){\footnotesize At parent};
\end{scope}
\begin{scope}[shift={(330,15.5)}]
\draw[thick] (0,0) -- ++(-2,0) -- ++(0,13) -- ++(2,0);
\draw[thick] (9,0) -- ++(2,0) -- ++(0,13) -- ++(-2,0);
v\fill[red] (0,2) rectangle +(4,4) ++(5,0) rectangle +(4,4) ++(0,5) rectangle +(4,4)++(-5,0) rectangle +(4,4);
\end{scope}
\end{tikzpicture}
\caption{Illustration of the one-stage ULV factorization process.}
\label{fig:ULV}
\end{figure}

Note that in the one-stage case presented above, we have to perform LQ factorizations of two matrices with order \(n\) rows and columns, assuming \(r\) is small; therefore the cost is \(\mathcal{O}(n^3)\). To bring the asymptotic cost down, we need more levels in the HSS tree. In the next step, we consider the two-stage HSS decomposition in Equation (\ref{eqn:2level_hss_nested}), i.e., a three-level tree. We assume that the two diagonal blocks (children 1 and 2) are already transformed into ULV form (\ref{eqn:1level_ulv}), via \(U = \text{diag}(\Omega_3^{-1}, \Omega_4^{-1}, \Omega_5^{-1}, \Omega_6^{-1})\) and \(V = \text{diag}(Q_3, Q_4, Q_5, Q_6)\). The remaining uneliminated blocks are 
\begin{equation}
 D_1 = \begin{bmatrix} W_{3;b}Q_{3;b}^* & B_{3,4}V_4^*Q_{4;b}^* \\ 
  B_{4,3}V_3^*Q_{3;b}^*  & W_{4;b}Q_{4;b}^* \end{bmatrix} , \ 
D_2 = \begin{bmatrix} W_{5;b}Q_{5;b}^* & B_{5,6}V_6^*Q_{6;b}^* \\ 
  B_{6,5}V_5^*Q_{5;b}^*  & W_{6;b}Q_{6;b}^* \end{bmatrix} .
\label{eqn:d1d2}
\end{equation}
The two transformations \(\text{diag}(\Omega_3, \Omega_4)\) and \(\text{diag}(\Omega_5, \Omega_6)\) can be respectively applied to the off-diagonal blocks (1,2) and (2,1) of the matrix \(A\). Due to the nested basis property (see (\ref{eqn:nested_basis})), \(U_3^{\text{big}}\) and \(U_4^{\text{big}}\) are already annihilated to \(\begin{bmatrix}0 \\ I\end{bmatrix}\). Therefore, the only nonzero part of the (1,2) block of \(A\) is \(\setlength{\arraycolsep}{-0.2em}U_1 B_{1,2}V_2^* \begin{bmatrix}V_5^*Q_{5;t}^*&&V_5^*Q_{5;b}^*\\&V_6^*Q_{6;t}^*&&V_6^*Q_{6;b}^*\end{bmatrix} \); similarly for the (2,1) block of \(A\).

At the parent nodes 1 and 2, which are non-root nodes and have \(U_1\) and \(U_2\) bases associated with them,  we use the transformations \(\Omega_1 U_1 = \begin{bmatrix} 0 \\ I \end{bmatrix}\) and \(\Omega_2 U_2 = \begin{bmatrix} 0 \\ I \end{bmatrix}\) to introduce the zero blocks. Then, we apply the above annihilation and transformation to the diagonal blocks \(D_1\) and \(D_2\) (see (\ref{eqn:d1d2})), followed by the \(LQ\) decomposition of each top part. Eventually, at the root node, only \(\mathcal{O}(r)\) unknowns are left and a regular \(LU\) factorization (with pivoting if needed) is used.

The two-stage transformation process can be written as follows:
{\scriptsize
\newlength{\myseparator}
\newlength{\mycol}
\begin{align}
&
\Gamma_{1;b\leftrightarrow 2;t}
\begingroup\setlength{\arraycolsep}{0.04em}
\begin{bmatrix} I & & & \\ & \Omega_1 & &  \\  & & I & \\ &&& \Omega_2 \end{bmatrix}\! 
\begin{bmatrix} \Gamma_{3;b\leftrightarrow 4;t}\! & \\ & \!\Gamma_{5;b\leftrightarrow 6;t}\end{bmatrix}\! 
\begin{bmatrix}\Omega_3 & & &  \\  & \Omega_4 & & \\ & & \Omega_5 & \\ &  & &\Omega_6\end{bmatrix}\!
A\! 
\begin{bmatrix} Q_3^* & & &  \\  & Q_4^* & & \\ & & Q_5^* & \\  & & & Q_6^*\end{bmatrix}\!
\begin{bmatrix} \Gamma_{3;b\leftrightarrow 4;t}^T\! & \\ & \!\Gamma_{5;b\leftrightarrow 6;t}^T\end{bmatrix}\!
\begin{bmatrix} I &&& \\ & Q_1^* && \\  && I & \\ &&& Q_2^* \end{bmatrix}
\endgroup
\Gamma_{1;b\leftrightarrow 2;t}^T   \nonumber \\ 
  &= 
\renewcommand{\arraystretch}{1.4}
\newcommand{\rst}[1]{\renewcommand{\arraystretch}{1.0}\setlength{\arraycolsep}{#1 em}}
\setlength{\arraycolsep}{0em}
\setlength{\myseparator}{0.8pt}
\setlength{\arrayrulewidth}{\myseparator}
\newcommand{\myrule}{\vrule width \myseparator}
\newcolumntype{C}{!{\myrule}c!{\myrule}}
\newcolumntype{L}{!{\myrule}c}
\newcolumntype{R}{c!{\myrule}}
\newcommand{\mcc}[1]{\multicolumn{1}{C}{#1}}
\newcommand{\mcl}[1]{\multicolumn{1}{L}{#1}}
\newcommand{\mcr}[1]{\multicolumn{1}{R}{#1}}
\newcommand{\mystrut}{\rule[-1.1em]{0pt}{2.7em}}
\newcommand{\Aoneone}{\mcc{(\Omega_1 L_{4,3})_b}}
\newcommand{\Aonetwo}{\mcr{(\Omega_1 L_{3,4})_b}}
\newcommand{\Aonethree}{\mcr{W_{1;b} Q^*_{1;t}}}
\newcommand{\Aonefourandfiveandsix}{\multicolumn{3}{R}{\mystrut\rst{-0.2} B_{1,2}V_2^*\begin{bmatrix}V_5^*Q_{5;t}^*&&V_5^*Q_{5;b}^*\\&V_6^*Q_{6;t}^*&&V_6^*Q_{6;b}^*\end{bmatrix}\rst{0.2}\begin{bmatrix}I \\ & Q_2^*\end{bmatrix}}}
\newcommand{\Aoneandtwoseven}{\mcr{\multirow{2}{*}{\rule{0pt}{2em}\(D_0\)}}}
\newcommand{\Atwooneandtwoandthree}{\multicolumn{3}{C}{\mystrut\rst{-0.2} B_{2,1}V_1^*\begin{bmatrix}V_3^*Q_{3;t}^*&&V_3^*Q_{3;b}^*\\&V_4^*Q_{4;t}^*&&V_4^*Q_{4;b}^*\end{bmatrix}\rst{0.2}\begin{bmatrix}I \\ & Q_1^*\end{bmatrix}}}
\newcommand{\Atwofour}{\mcr{(\Omega_2 L_{6,5})_b}}
\newcommand{\Atwofive}{\mcr{(\Omega_2 L_{5,6})_b}}
\newcommand{\Atwosix}{\mcr{W_{2;b} Q^*_{2;t}}}
\newcommand{\Adummy}{\mcr{}}
\setlength{\mycol}{6.7em}
\!\left[\,\begin{array}{p{\mycol}p{\mycol}p{\mycol}p{\mycol}p{\mycol}p{\mycol}p{3.7em}}
\\[-0.8em]
\cline{1-1}
\mcc{L_3}              &                        &            &                        &                        &           &                  \\ 
\cline{1-2}
\mcc{0}                 &              \mcr{L_4} &            &                        &                        &           &                  \\ 
\cline{1-3}
\mcc{(\Omega_1 L_{4,3})_t} & \mcr{(\Omega_1 L_{3,4})_t} & \mcr{L_1} &                        &                        &           &                  \\ 
\cline{1-4}
\multicolumn{3}{C}{}                                         &              \mcr{L_5} &                        &           &                  \\ 
\cline{4-5}
\multicolumn{3}{C}{0}                                         &                 \mcr{0} &              \mcr{L_6} &           &                  \\ 
\cline{4-6}
\multicolumn{3}{C}{}                                         & \mcr{(\Omega_2 L_{6,5})_t} & \mcr{(\Omega_2 L_{5,6})_t} & \mcr{L_2} &                  \\ 
\hline
\Aoneone               &               \Aonetwo & \Aonethree &                                \Aonefourandfiveandsix & \Aoneandtwoseven \\     
\cline{1-6}
\Atwooneandtwoandthree                                       &              \Atwofour &        \Atwofive &  \Atwosix &          \Adummy \\     
\hline\\[-0.9em]
\end{array}\,\right]
\label{eqn:2level_ulv_transform}
\end{align}
}

The algorithm is presented in Algorithm~\ref{alg:ULV}. The complexity is \(\mathcal{O}(r^2 n)\)~\cite{chandrasekaran2006fast,xia2013randomized}. Notice that the output of the algorithm is, at each non-root node \(\tau\), the \(Q_\tau\) and \(L_\tau\) matrices that represent the ULV factors, but also the matrix \(W_\tau=\Omega_\tau D_\tau\) and the matrix \(\tilde{V}_\tau\), that accumulates the actions of \(V\) bases and the \(Q\) transformations, as shown in lines~\ref{algline:vhat} and~\ref{algline:vtilde} of the algorithm. \(\tilde{V}_\tau\) is conceptually similar to \(V_\tau^{\text{big}}\), except it has only \(\mathcal{O}(r)\) rows, corresponding to the uneliminated variables. The matrices \(W_\tau\) and \(\tilde{V}_\tau\) are useful for the solution phase, as shown in the next section.

\begin{algorithm}[!ht]
\KwData{HSS form: \(D_\tau\) (leaves), \(U_\tau\), \(V_\tau\) (all nodes except root), \(B_{\nu_1, \nu_2}\) and \(B_{\nu_2, \nu_1}\) (non-leaves).}
\KwResult{ULV factors: \(Q_\tau\) orthonormal, \(L_\tau\) lower triangular (all nodes except root). \(LU\) at root.\newline
\(W_\tau\) and \(\tilde{V}_\tau\) to be used in solution step.
}
\BlankLine
\ForEach{node \(\tau\) in topological order (fine to coarse)}{
  \eIf{node \(\tau\) is a non-leaf}{
    \(D_\tau = \begin{bmatrix} \tilde{D}_{\nu_1} & B_{\nu_1,\nu_2} \tilde{V}_{\nu_2;b}^* \\ B_{\nu_2,\nu_1} \tilde{V}_{\nu_1;b}^* & \tilde{D}_{\nu_2} \end{bmatrix}\)\\
    \If{node \(\tau\) is not the root node}{
      \(\hat{V}_\tau = \begin{bmatrix} \tilde{V}_{\nu_1;b} & 0 \\ 0 & \tilde{V}_{\nu_2;b} \end{bmatrix} V_\tau\)~\label{algline:vhat}\\
    }
  }{
      \(\hat{V}_\tau = V_\tau\)
  }
  \eIf{node \(\tau\) is the root node}{
    \(\begin{bmatrix} P_\tau, L_\tau, U_\tau \end{bmatrix} = \text{LU}\left( D_{\tau} \right)\)
  }{
    \(W_\tau = \Omega_\tau D_\tau = \begin{bmatrix} -E_\tau^r & I \\ I & 0 \end{bmatrix}
        {\Pi^r_\tau}^T D_\tau
    = \begin{bmatrix} W_{\tau;t} \\ W_{\tau;b}  \end{bmatrix} \)\\
    \(\text{LQ}\left( W_{\tau;t} \right) = \begin{bmatrix} L_\tau & 0 \end{bmatrix} \begin{bmatrix} Q_{\tau;t} \\ Q_{\tau;b} \end{bmatrix}\)\\
    \(\tilde{V}_\tau = Q_{\tau} \hat{V}_\tau = \begin{bmatrix} \tilde{V}_{\tau;t} \\ \tilde{V}_{\tau;b} \end{bmatrix}\)~\label{algline:vtilde}\\
    \(\tilde{D}_{\tau} = W_{\tau;b} Q_{\tau;b}^*\)
  }
}
\caption{ULV-like factorization of a non-symmetric matrix in HSS form.}
\label{alg:ULV}
\end{algorithm}

\subsection{Solution using ULV factorization}\label{sec:solve}
The ULV-factored form (\ref{eqn:2level_ulv_transform}) can be used to solve a linear system \(Ax=b\). We still use the two-stage HSS example (three-level tree) to explain the solution procedure.

Consider a partitioning of the right-hand side \(b\) and the solution vector \(x\) along the cluster tree: \(b=\begin{bmatrix}b(I_1,:)\\b(I_2,:)\end{bmatrix}=\begin{bmatrix}b_1\\b_2\end{bmatrix}\) and similarly, \(x=\begin{bmatrix}x_1\\x_2\end{bmatrix}\) and the one-stage ULV factorization given in Equation (\ref{eqn:1level_ulv_transform}). The solution \(x\) can be obtained by the following five steps:
\begin{enumerate}
\item Transform the right-hand side: \(\tilde{b}_1 = \Omega_1 b_1\), and 
  \(\tilde{b}_2 = \Omega_2 b_2\);
\item Forward substitution: \(y_1 = L_1^{-1} \tilde{b}_{1;t}\),\ 
  \(y_2 = L_2^{-1} \tilde{b}_{2;t}\);
\item Update right-hand side:\\
  \( b_{1;b} = \tilde{b}_{1;b} - W_{1;b} Q^*_{1;t}\: y_1 - B_{1,2}V_2^* Q^*_{2;t}\: y_2\),\\
  \( b_{2;b} = \tilde{b}_{2;b} - B_{2,1}V_1^* Q^*_{1;t}\: y_1 - W_{2;b} Q^*_{2;t}\: y_2 \);
\item Triangular solution at root: 
  \( x_0 = U_0^{-1} L_0^{-1}P_0\; \begin{bmatrix}b_1\\b_2\end{bmatrix}\).
\item Orthogonally transform back to the original solution:
  \( x_1 = Q^*_1\begin{bmatrix} y_1 \\ x_{0;t}\end{bmatrix}\), 
  \( x_2 = Q^*_2\begin{bmatrix} y_2 \\ x_{0;b}\end{bmatrix}\).
\end{enumerate}

Next, consider the two-stage ULV transformation given in Equation (\ref{eqn:2level_ulv_transform}). Algorithm~\ref{alg:solve} shows the complete procedure, which follows a bottom-up traversal of the HSS tree. We first apply all the transformations involving \(\Omega\)'s to the right-hand side \(b\), to obtain \(\tilde{b}\) (line~\ref{algline:btilde} in the Algorithm.) Then we obtain all the intermediate variable \(y_\tau\) for the non-root node \(\tau\) via forward substitution (line~\ref{algline:ytrs} in the Algorithm). Now looking at the last block row of (\ref{eqn:2level_ulv_transform}) involving \(D_0\), we need the contributions from the children of the root node (nodes 1 and 2). For example, the intermediate solution \(y_1\) coming from node 1 contributes to the terms \(W_{1;b} Q_{1;t}^* y_1\) and \(B_{2,1}V_1^* Q^*_{1;t} y_1\). Furthermore, there are contibutions coming from the grand children of the root node, i.e., nodes 3, 4, 5, and 6. For example, nodes 3 and 4 contribute via the term \(B_{2,1}V_1^*\begin{bmatrix}V_3^*Q_{3;t}^*&&V_3^*Q_{3;b}^*\\&V_4^*Q_{4;t}^*&&V_4^*Q_{4;b}^*\end{bmatrix}\begin{bmatrix}I \\ & Q_1^*\end{bmatrix}\begin{bmatrix}y_3\\y_4\\y_1\end{bmatrix}\). In the general case (arbitrary number of levels), \(b_0\) (updated right-hand side at root node) receives contributions from all the nodes in the tree, because the last block row of the \(L\) is full. In the algorithm, we accumulate these updates when going up the tree, as shown in lines~\ref{algline:btilde} and~\ref{algline:zupdate} of the algorithm. We illustrate this in more detail in Appendix~\ref{app:solve}.

Finally, the intermediate solution involving \(y\) needs to be transformed back to the original solution \(x\) (line~\ref{algline:xfromy} in the Algorithm). The complexity of Algorithm~\ref{alg:solve} is \(\mathcal{O}(rn)\)~\cite{chandrasekaran2006fast,xia2013randomized}.

\begin{algorithm}[!ht]
\KwData{ULV factors: \(Q_\tau\) orthonormal, \(L_\tau\) lower triangular (all nodes except root). \(LU\) at root.}
\KwResult{\(x\), solution of \(A x = b\).}
\BlankLine
\ForEach{node \(\tau\) in topological order (bottom-up traversal)}{
  \eIf{node \(\tau\) is a non-leaf}{
    \(b_\tau = \begin{bmatrix} \tilde{b}_{\nu_1;b} - W_{\nu_1;b} Q_{\nu_1;t}^* \: y_{\nu_1}
      - B_{\nu_1,\nu_2} \: z_{\nu_2} \\
                               \tilde{b}_{\nu_2;b} -           B_{\nu_2,\nu_1}\: z_{\nu_1}
                    - W_{\nu_2;b} Q_{\nu_2;t}^* \: y_{\nu_2} \end{bmatrix}\)
  }{
    \(b_\tau=b(I_\tau,:)\)
  }
  \eIf{node \(\tau\) is the root node}{
    \(x_\tau = U_\tau^{-1} L_\tau^{-1} P_\tau b_\tau = \begin{bmatrix} x_{\tau;t} \\ x_{\tau;b} \end{bmatrix}\)
  }{
    \(\tilde{b}_\tau = \Omega_\tau b_\tau = \begin{bmatrix} -E_\tau^r & I \\ I & 0 \end{bmatrix} {\Pi^r_\tau}^T b_\tau
    = \begin{bmatrix} \tilde{b}_{\tau;t} \\ \tilde{b}_{\tau;b} \end{bmatrix}\)\label{algline:btilde}\\
    \(y_\tau = L_\tau^{-1} \tilde{b}_{\tau;t}\)\label{algline:ytrs}\\
    \eIf{node \(\tau\) is a non-leaf}{
      \(z_\tau = V_\tau^* \begin{bmatrix} z_{\nu_1} \\ z_{\nu_2} \end{bmatrix} + \tilde{V}^*_{\tau;t}\: y_\tau\)\label{algline:zupdate}
    }{
      \(z_\tau = \tilde{V}^*_{\tau;t}\: y_\tau\)
    }
  }
}
\ForEach{node \(\tau\) in reverse topological order (top-down traversal)}{
  \eIf{node \(\tau\) is a non-leaf}{
    \(x_{\nu_1} = Q^*_{\nu_1} \begin{bmatrix} y_{\nu_1} \\ x_{\tau;t} \end{bmatrix}\) , \quad
    \(x_{\nu_2} = Q^*_{\nu_2} \begin{bmatrix} y_{\nu_2} \\ x_{\tau;b} \end{bmatrix}\)\label{algline:xfromy}
  }{
    \(x(I_\tau,:)=x_\tau\)
  }
  
}
\caption{Solution of a linear system \(Ax=b\) after ULV-like factorization, for a non-symmetric matrix.}
\label{alg:solve}
\end{algorithm}

\section{Distributed-memory parallelism}
\label{sec:para}

In this section, we present our distributed-memory algorithms. We mostly focus on the implementation of the HSS compression algorithm, as this is the most complicated of all HSS operations but also the most critical for performance. In Section~\ref{sec:adaptive}, we present a novel parallel adaptive sampling mechanism.

\subsection{Task mapping}
\label{sec:mapping}

The HSS tree presented in Section~\ref{sec:HSSdef} is a task graph and data-dependency graph for all the different operations: compression, factorization, solution, and product. The tree structure allows for two levels of parallelism. \emph{Tree parallelism} comes from the fact that nodes lying on different branches of the tree can be processed in parallel, independently of one another. \emph{Node parallelism} consists in assigning a node of the tree to multiple processes. We enforce node parallelism by using parallel kernels from PBLAS~\cite{choi1996proposal} and ScaLAPACK~\cite{blackford1997scalapack}.

We rely on a static mapping technique to assign tasks to different processes. We use the idea of the \emph{proportional mapping} by Pothen and Sun~\cite{pothen1993map}, which is popular for mapping tasks along the elimination tree of sparse factorizations. The mapping process consists in a top-down traversal of the tree. All the processes are assigned to work on the root node, because this is the last task to be executed during a bottom-up traversal (e.g., compression, factorization) and the first task to be executed during a top-down traversal (e.g., matrix-vector product and triangular solution). Then, for every node in the tree, the list of processes working at that node is split among its children, proportionally to the weights (determined according to a given metric) of the subtrees rooted at these children. Consider a parent node \(f\) in the tree with \(\mathit{nc}_f\) children. Let \(p_f\) be the number of processes working at that node and \(W_i\) be the load of the subtree rooted at a child \(i\). The number of processes given to node \(i\) is
\[p_i=\frac{W_i}{\sum_{j=1}^{\mathit{nc}_f}W_j}\cdot p_f\]
This procedure is applied in a recursive fashion to all the children of \(f\); the recursion stops when leaf nodes are reached or entire subtrees are mapped onto single processes, which happens because the number of nodes in the tree is commonly much larger than the number of processes.

The usual metric used at each step of the mapping is the workload of each subtree. However, in our case, we cannot use this because we do not know in advance the cost of processing a node since it depends on the ranks found at that node. Instead, we use the size of the interval \(I_\tau\) associated with each node \(\tau\). The idea is that, at leaf nodes, the compression cost (computing local samples and performing Interpolative Decomposition) is proportional to the size of the interval. If the ranks found at different branches of the tree are balanced, workloads will be balanced. Otherwise, workloads might be unbalanced, leading to poorer performance of the compression process. However, after the compression is done, the tree can be remapped using the rank information, which can be useful for subsequent operations (factorization, etc.) or for improving compression times for different problems from the same application. We illustrate this in Section~\ref{sec:comb}.

An interesting property of the proportional mapping is that the traversal of every process (i.e., the set of tasks that this process executes and the order in which those are processed) is fully known in advance. Indeed, every process is in charge of a sequential subtree and takes part in the computation of the parallel nodes in the path between that subtree and the root of the elimination tree; this defines a single possible traversal. Denoting by \(i\) the root of the sequential subtree mapped on a given process, the traversal followed by that process consists of a postorder traversal of the subtree rooted at \(i\) followed by the path from \(i\) to the root node. This makes the code easier to write.

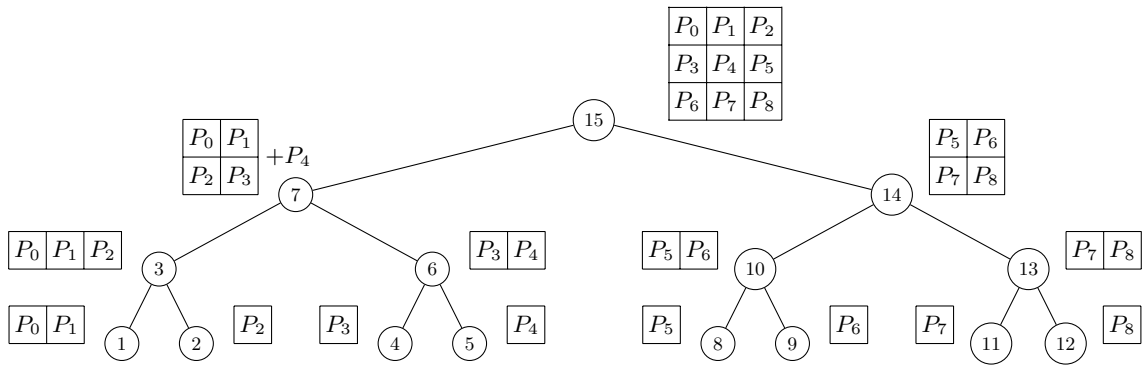
\begin{figure}[!ht]
\centering
\tikzstyle{cnode} = [draw, circle,scale=0.8]
\tikzstyle{level 1} = [level distance=.06\textwidth, sibling distance=.48\textwidth]
\tikzstyle{level 2} = [sibling distance=.22\textwidth]
\tikzstyle{level 3} = [sibling distance=.06\textwidth]
\begin{tikzpicture}\small
  \node[cnode](15){15}
  child{node[cnode](7){7}
    child{node[cnode](3){3}
      child{node[cnode](1){1}}
      child{node[cnode](2){2}}
    }
    child{node[cnode](6){6}
      child{node[cnode](4){4}}
      child{node[cnode](5){5}}
    }
  }
  child{node[cnode](14){14}
    child{node[cnode](10){10}
      child{node[cnode](8){8}}
      child{node[cnode](9){9}}
    }
    child{node[cnode](13){13}
      child{node[cnode](11){11}}
      child{node[cnode](12){12}}
    }
  };
\draw[step=.5,shift=(15)] (.99,0) grid +(1.51,1.51) +(0.25,1.25)node{\(P_0\)} +(0.75,1.25)node{\(P_1\)} +(1.25,1.25)node{\(P_2\)} +(0.25,0.75)node{\(P_3\)} +(0.75,0.75)node{\(P_4\)} +(1.25,0.75)node{\(P_5\)} +(0.25,0.25)node{\(P_6\)} +(0.75,0.25)node{\(P_7\)} +(1.25,0.25)node{\(P_8\)};
\draw[step=.5,shift=(7)] (-1.5,0) grid +(1,1) +(0.25,0.75)node{\(P_0\)} +(0.75,0.75)node{\(P_1\)} +(0.25,0.25)node{\(P_2\)} +(0.75,0.25)node{\(P_3\)} +(1.4,0.51) node{+\(P_4\)};
\draw[step=.5,shift=(3)] (-2,0) grid +(1.5,0.5) +(0.25,0.25)node{\(P_0\)} +(0.75,0.25)node{\(P_1\)} +(1.25,0.25)node{\(P_2\)};
\draw[step=.5,shift=(1)] (-1.5,0) grid +(1.,0.5) +(0.25,0.25)node{\(P_0\)} +(0.75,0.25)node{\(P_1\)};
\draw[step=.5,shift=(2)] (.5,0) grid +(0.5,0.5) +(0.25,0.25)node{\(P_2\)};
\draw[step=.5,shift=(6)] (.5,0) grid +(1.,0.5) +(0.25,0.25)node{\(P_3\)} +(0.75,0.25)node{\(P_4\)};
\draw[step=.5,shift=(4)] (-1,0) grid +(0.5,0.5) +(0.25,0.25)node{\(P_3\)};
\draw[step=.5,shift=(5)] (.5,0) grid +(0.5,0.5) +(0.25,0.25)node{\(P_4\)};
\draw[step=.5,shift=(14)] (.5,0) grid +(1,1) +(0.25,0.75)node{\(P_5\)} +(0.75,0.75)node{\(P_6\)} +(0.25,0.25)node{\(P_7\)} +(0.75,0.25)node{\(P_8\)};
\draw[step=.5,shift=(10)] (-1.5,0) grid +(1.,0.5) +(0.25,0.25)node{\(P_5\)} +(0.75,0.25)node{\(P_6\)};
\draw[step=.5,shift=(8)] (-1,0) grid +(0.5,0.5) +(0.25,0.25)node{\(P_5\)};
\draw[step=.5,shift=(9)] (.5,0) grid +(0.5,0.5) +(0.25,0.25)node{\(P_6\)};
\draw[step=.5,shift=(13)] (.5,0) grid +(1.,0.5) +(0.25,0.25)node{\(P_7\)} +(0.75,0.25)node{\(P_8\)};
\draw[step=.5,shift=(11)] (-1,0) grid +(0.5,0.5) +(0.25,0.25)node{\(P_7\)};
\draw[step=.5,shift=(12)] (.5,0) grid +(0.5,0.5) +(0.25,0.25)node{\(P_8\)};
\end{tikzpicture}
\caption{Proportional mapping of an HSS tree with 9 processes and uniform weights. Every node is associated with a 2D grid of processes and, sometimes, a few idle processes.}
\label{fig:mapping}
\end{figure}

As we just saw, a node of the HSS tree can be mapped onto several processes. Within a node, our choice is to perform all the arithmetic operations with PBLAS and ScaLAPACK. All the matrices that we handle are distributed following a 2D block-cyclic scheme and each node of the HSS tree is associated with a 2D grid of processes that handle the computations. We typically try to make the grid as square as possible, as advised in the ScaLAPACK documentation~\cite{blackford1997scalapack}, but the code can accommodate any kind for grid. For example, if 32 processes work at a node, our grid has \(\left\lfloor\sqrt{32}\right\rfloor=5\) rows and \(\lfloor32/5\rfloor=6\) columns. In this example, \(32-6\times5=2\) processes stay idle at that node. However, it does not mean these processes are idle throughout the whole computation; they are idle at that node, but can be active at ancestors or descendants of that node. We illustrate this situation in Figure~\ref{fig:mapping}. In this example, node 7 is mapped on processes \(P_0\) to \(P_4\), but \(P_4\) is out of the 2D grid associated with node 7 and is thus idle at that node. However it is active at nodes 4 and 6 (descendants of 7) and 15 (ancestor of 7). At a given node mapped on \(P\) processes, the associated grid is \(P_r\times P_c\) and there are at most \(\lfloor\sqrt{P}\rfloor-1\) idle processes.

\subsection{Parallel compression}
\label{sec:parcompr}

We provide some details about our implementation of the parallel HSS construction (compression) algorithm. The first stage of the compression algorithm is to generate random vectors. In STRUMPACK, different generators can be used: the legacy \texttt{rand} C function, or advanced generators from the \texttt{C++11} standard, like the Mersenne Twister~\cite{matsumoto1998mersenne}. They can be combined with a postprocessing that enforces certain distribution of random numbers, e.g., uniform or normal.

The second stage is to generate the samples \(S^r = AR^r\) and \(S^c = A^* R^c\). For this, the compression algorithm needs either access to a user-given matrix-vector product or explicit access to the whole matrix \(A\). We require the input matrix to be distributed in 2D block-cyclic form and we use the PBLAS matrix-matrix product \texttt{PxGEMM} to compute the product.

The third stage is a topological traversal of the tree, where at each node, a local sample is formed then compressed and updated. To form the sample we need access to some selected elements of the matrix. For this, the compression algorithm needs either access to a user-given routine that provides selected elements or explicit access to the whole matrix \(A\). If the input matrix \(A\) is explicitly given (in 2D block-cyclic form), we distribute it so that at each stage of the compression, a process can extract selected elements without communicating with processes working at other nodes of the HSS tree. This is done by traversing the tree following a serialized postorder (i.e., all the processes traverse the whole HSS tree synchronously). At each node, a piece of the original matrix (shared by all the processes) is redistributed to the subset of processes that work at that node. The diagonal blocks of \(A\) correspond to leaves of the HSS tree and are redistributed to the processes working at these nodes, so that they can extract a diagonal block \(D_\tau\) without communicating with processes mapped at other nodes. Similarly, the off-diagonal blocks are also distributed along the mapping of the tree, so that the \(B_{\nu_1,\nu_2}\) and \(B_{\nu_2,\nu_1}\) matrices at non-leaf nodes can be extracted without communication. For each block, we rely on a 2D block-cyclic distribution using the process grid associated with the corresponding node.
  
We provide an example in Figure~\ref{fig:dist}, corresponding to the mapping in Figure~\ref{fig:mapping}. Consider node 1. The first step in the compression at node 1 is to extract \(D_1=A(I_1,I_1)\) from the input matrix. These entries are distributed on \(P_0\) and \(P_1\) and readily available. Then, at node 3, which is mapped on \(P_0\), \(P_1\) and \(P_2\), matrix \(B_{1,2}=A(I^r_1,I^c_2)\) is extracted by selecting some rows and columns of \(A(I_1,I_2)\). \(A(I_1,I_2)\) is distributed on \(P_0\), \(P_1\) and \(P_2\), therefore the extraction can be done without communicating with processes working at other nodes.

\begin{figure}[!ht]
\centering
\begin{tikzpicture}[scale=.92]\footnotesize
\draw (0,0) -- +(1,0) -- +(1,1) -- +(0,1) -- cycle;
\node at (.5,.5){\(P_0\,\,P_1\)};
\foreach \i in {1,...,7} {
  \draw (\i,-\i) -- +(1,0) -- +(1,1) -- +(0,1) -- cycle;
  \pgfmathtruncatemacro{\ione}{\i+1};
  \node at (\i+.5,-\i+.5){\(P_{\ione}\)};
}
\foreach \i in {1,3,5,7} {
  \draw (\i-1,-\i) -- +(1,0) -- +(1,1) -- +(0,1) -- cycle;
  \draw (\i,-\i+1) -- +(1,0) -- +(1,1) -- +(0,1) -- cycle;
}
\foreach \i in {3,5,7} {
  \pgfmathtruncatemacro{\ione}{\i+1};
  \node at (\i+.5,-\i+1.5){\(P_\i\,\,P_\ione\)};
  \node at (\i-.5,-\i+0.5){\(P_\i\,\,P_\ione\)};
}
\node at (1.5,.5){\(P_0\!P_1\!P_2\)};
\node at (0.5,-.5){\(P_0\!P_1\!P_2\)};
\foreach \i in {1,5} {
  \draw (\i+1,-\i) -- +(2,0) -- +(2,2) -- +(0,2) -- cycle;
  \draw (\i-1,-\i-2) -- +(2,0) -- +(2,2) -- +(0,2) -- cycle;
}
\draw[step=.5,shift={(2.1,-0.5)}] +(0.25,0.75)node{\(P_0\)} +(0.75,0.75)node{\(P_1\)} +(0.25,0.25)node{\(P_2\)} +(0.75,0.25)node{\(P_3\)} +(1.4,0.51) node{+\(P_4\)};
\draw[step=.5,shift={(0.1,-2.5)}] +(0.25,0.75)node{\(P_0\)} +(0.75,0.75)node{\(P_1\)} +(0.25,0.25)node{\(P_2\)} +(0.75,0.25)node{\(P_3\)} +(1.4,0.51) node{+\(P_4\)};
\draw[step=.5,shift={(6.5,-4.5)}] +(0.25,0.75)node{\(P_5\)} +(0.75,0.75)node{\(P_6\)} +(0.25,0.25)node{\(P_7\)} +(0.75,0.25)node{\(P_8\)};
\draw[step=.5,shift={(4.5,-6.5)}] +(0.25,0.75)node{\(P_5\)} +(0.75,0.75)node{\(P_6\)} +(0.25,0.25)node{\(P_7\)} +(0.75,0.25)node{\(P_8\)};
\draw (4,-3) -- +(4,0) -- +(4,4) -- +(0,4) -- cycle;
\draw[step=.5,shift={(5.25,-1.75)}] +(0.25,1.25)node{\(P_0\)} +(0.75,1.25)node{\(P_1\)} +(1.25,1.25)node{\(P_2\)} +(0.25,0.75)node{\(P_3\)} +(0.75,0.75)node{\(P_4\)} +(1.25,0.75)node{\(P_5\)} +(0.25,0.25)node{\(P_6\)} +(0.75,0.25)node{\(P_7\)} +(1.25,0.25)node{\(P_8\)};
\draw (0,-7) -- +(4,0) -- +(4,4) -- +(0,4) -- cycle;
\draw[step=.5,shift={(1.25,-5.75)}] +(0.25,1.25)node{\(P_0\)} +(0.75,1.25)node{\(P_1\)} +(1.25,1.25)node{\(P_2\)} +(0.25,0.75)node{\(P_3\)} +(0.75,0.75)node{\(P_4\)} +(1.25,0.75)node{\(P_5\)} +(0.25,0.25)node{\(P_6\)} +(0.75,0.25)node{\(P_7\)} +(1.25,0.25)node{\(P_8\)};
\end{tikzpicture}
\caption{Distribution of the input matrix conforming to the mapping in
 Figure~\ref{fig:mapping}.}
\label{fig:dist}
\end{figure}

After building random vectors, computing samples, and distributing the input matrix, the postorder traversal starts. Serial subtrees (subtrees mapped on one process) are processed by a sequential compression routine that relies on BLAS and LAPACK kernels (which is usually better than using PBLAS or ScaLAPACK kernels serially). Then, parallel nodes are processed using PBLAS and ScaLAPACK operations. The main computational kernels are matrix-matrix products (performed with PBLAS \texttt{PxGEMM}) and the Interpolative Decomposition procedure described previously. For the latter, we explored two options:
\begin{enumerate}
\item Modifying the \texttt{xGEQP3} and \texttt{PxGEQPF} from LAPACK and ScaLAPACK respectively. These routines perform a QR factorization with column pivoting but they compute the full factorization. We modified them to embed our compression threshold \(\varepsilon\). The factorization stops when the norm of the pivot column becomes too small, i.e., \(\frac{R_{ii}}{R_{11}}\leq\varepsilon\), with \(R\) the partial \(R\) factor. The number of columns actually eliminated is the \(\varepsilon\)-rank of the block to be compressed.
\item Implementing a Modified Gram-Schmidt (MGS) algorithm with column pivoting. The parallel implementation uses 2D block-cyclic operations. A similar version was used in Hsolver~\cite{wang2013efficient}.
\end{enumerate}
In a parallel setting, we have not observed much difference in performance between the two options. In a serial setting, the modified xGEQP3 routine, which uses a BLAS3 implementation, is typically two to three times faster than our BLAS2 MGS implementation.

\subsection{Adaptive sampling mechanism}
\label{sec:adaptive}

The algorithm in Section~\ref{sec:compr} assumes that the HSS rank \(r\) of the input matrix is known, so that the number of sample vectors \(d\) (number of columns of \(R^r\) and \(R^c\)) is chosen to be a tight upper bound of \(r\). Indeed, \(d\) needs to be larger than \(r\) to get a stable compression, but it also needs to be not too large, because the sampling process requires \(\mathcal{O}(dn^2)\) operations and can dominate the other parts of the compression stage.

In practice, \(r\) is rarely known. For some specific applications, we have a rough idea of its value, as described in Section~\ref{sec:compr}. In order to get a more black-box compression process, it is important to design an \emph{adaptive sampling} mechanism. This is mentioned in~\cite{martinsson2011fast,xia2012superfast} but neither an algorithm nor an implementation is described in detail. Here we explain our parallel adaptive sampling algorithm and implementation. The idea is to start with a low number of random vectors \(d\), and whenever the rank found during Interpolative Decomposition is too large, \(d\) is increased. Instead of restarting the compression from scratch, we keep the generators that have been computed and the computation restarts at the node(s) where the rank was too large.

In a serial setting, the sketch of the algorithm is the following. When the rank at a given node \(\tau_{fail}\) is too large, add new columns to \(R^r\) and \(R^c\), compute the new columns of \(S^r\) and \(S^c\) with a product, and restart the postorder traversal:
\begin{enumerate}
\item At nodes preceding \(\tau_{fail}\), keep the generators (\(D\), \(U\), etc.) that were previously computed. Update \(S^r_{loc}\) and \(S^c_{loc}\) with new columns.
\item At \(\tau_{fail}\), update \(S^r_{loc}\) and \(S^c_{loc}\) with new columns, and recompute the Interpolative Decomposition. If the rank is again too large, restart again, otherwise proceed to the next node.
\item At nodes following \(\tau_{fail}\), proceed as before.
\end{enumerate}
In this serial mechanism, a node can have three states: it can be \untouched{} if it has never been traversed before, \partially{} if the local samples have been computed but the rank obtained by Interpolative Decomposition was found too high (i.e., the traversal restarts because of this node), or \compressed{} if the generators have been successfully computed. There can be at most one \partially{} node in the tree. All the nodes that precede that node in the postorder are necessarily \compressed{}, and all the nodes that follow that node in the postorder are necessarily \untouched{}.

In a parallel execution, since we follow a parallel topological ordering of the tree instead of a serial postorder, we have different options. The choice we made is to implement a ``late notification'' mechanism. Whenever a process finds that the number of random vectors is not sufficient, it does not immediately notify the other processes. Instead, it simply invalidates
the current node by leaving it \untouched{}. Then, whenever a parent node is activated, we check the state of its two children. If they are not both \compressed{}, the parent node is left \untouched{}. Therefore, all the ancestors of the node that failed are left untouched. All the processes meet at the root node and can generate new random vectors, recompute samples, and restart the traversal. The difference with the serial case is that the tree can contain several \partially{} nodes. All the descendants of these \partially{} (i.e., failed) nodes are compressed, and all their ancestors have been left \untouched{}. This adaptive sampling mechanism is shown in Algorithm~\ref{alg:adapt}.

The main idea of this approach is that the different branches make as much progress as possible as long as the number of random vectors is sufficient. In the serial case, whenever a node fails, the traversal restarts with more random vectors, meaning that the subsequent branches will be processed with more -- and maybe unnecessary -- random vectors. As a consequence, different executions on different numbers of processes will lead to slightly different ranks and HSS representations.

Another choice, that we have not implemented, would be an ``early notification'' mechanism where processes are notified as early as possible that a node has failed somewhere in the tree. This is more complicated to implement and requires asynchronous communications to avoid barriers at each level or node of the tree. It is not clear that it would be significantly faster.

\begin{algorithm}[!ht]
\eIf{myid is in the 2D grid of \(\tau\)}{
  \If{\(\tau\) non-leaf and not all children are \compressed{}}{
    state stays \untouched{}\\
    \textbf{return}
  }
  \tcp{Sampling}
  \eIf{node is \untouched{}}{
    Extract \(D\) or \(B_{12}, B_{21}\), compute local samples \(S^r_{loc}\) and \(S^c_{loc}\)
  }{
    Compute updates to the samples, e.g., \(S^r_{upd}-DR^r_{upd}\) or \(\begin{bmatrix}{S^r_{\nu_1}}_{upd}-B_{12} {R^r_{\nu_2}}_{upd} \\ {S^r_{\nu_2}}_{upd}-B_{21} {R^r_{\nu_1}}_{upd}\end{bmatrix}\)
  }
  \tcp{Interpolative Decomposition}
  \If{node is \partially{}}{
    \tcp{Merge updates into the samples and random vectors}
    \(R^r\gets\left[R^r\, R^r_{upd}\right],\ S^r\gets\left[S^r\, S^r_{upd}\right],\ R^c\gets\left[R^c\, R^c_{upd}\right],\ S^c\gets\left[S^c\, S^c_{upd}\right]\)
  }
  \eIf{node is not \compressed{}}{
    Try Interpolative Decomposition of \(S^r\)
    \If{rank too small}{
      Throw away \(U\) and \(I_r\)\\
      Mark node as \partially{}\\
      \textbf{return}
    }
    Same for \(S^c\)
    \(S^r\gets S^r(I_r,:),\ S^c\gets S^c(I_r,:)\)
  }{
    \(S^r_{upd}\gets S^r_{upd}(I_r,:),\ S^c_{upd}\gets S^c_{upd}(I_r,:)\)
  }
  \tcp{Update}
  \eIf{node is \untouched{}}{
   \(R^r=V^*\times\ldots,\ R^c=V^*\times\ldots\)
  }{
   \(R^r_{upd}=V^*\times\ldots,\ R^c_{upd}=V^*\times\ldots\)
  }
  \If{node is \compressed{} and parent is \untouched{}}{
    \tcp{Merge updates into the samples and random vectors}
    \(R^r\gets\left[R^r\, R^r_{upd}\right],\ S^r\gets\left[S^r\, S^r_{upd}\right],\ R^c\gets\left[R^c\, R^c_{upd}\right],\ S^c\gets\left[S^c\, S^c_{upd}\right]\)
  }
  Mark node as \compressed{}
}{
  \tcp{myid is out of the 2D grid}
  Receive state from \(P_\tau\)
  \eIf{state==\compressed{}}{
     Receive ranks and indices from \(P_\tau\).
  }{
     restart()
  }
}
\caption{Processing a node \(\tau\).}
\label{alg:adapt}
\end{algorithm}

\clearpage

\subsection{Communication analysis}
\label{sec:comm}

We briefly analyze the amount of communication of our parallel compression algorithm. The analysis is similar to the one we derived previously on non-randomized algorithms~\cite{wang2013efficient}. We consider that each node of the HSS tree has the same rank \(r\) for its \(U\) and \(V\) generators; for some applications, a specific rank pattern can be used instead, as it is sometimes done in the literature~\cite{xia2013efficient,xia2013randomized}, but this is not our goal here. We also consider that, at the leaf nodes, the diagonal blocks have size \(\mathcal{O}(r)\). Finally, we assume that the number of processes is a power of 2, and the HSS tree is a complete binary tree. The pair [\#messages, \#words] is used to count the number of messages and the number of words transferred during a given operation, typically along the critical path. For example, a broadcast of \(w\) words among \(p\) processes is modeled as \([\log p,w\log p]\). This assumes that the broadcast follows a tree-based implementation; there are \(\log p\) steps on the critical path (any branch of the tree) and \(w\) words are transferred at each step, yielding \(\log p\) messages and \(w\log p\) words.

We denote \(n\) the size of the matrix and \(p\) the total number of processes. The parallel compression algorithm has three main steps:
\begin{enumerate}
\item Matrix-matrix product to compute the samples. We use the \texttt{PxGEMM} routine from PBLAS that relies on the SUMMA algorithm~\cite{vandegeijn1997summa} and can be modeled, asymptotically, as \([r\log p,\frac{r n}{\sqrt{p}}]\). This relies on the fact that, when computing a product \(S=A R\), the \texttt{PxGEMM} routine selects an algorithm that reduces communication based on the size of the operands \(A\), \(S\), \(R\). In our case, matrix \(A\) is the largest operand, so \texttt{PxGEMM} chooses an algorithm that communicates only \(S\) and \(R\). The selection strategy is described in~\cite{gunnels1998flexible}.
\item Initial distribution of the matrix along the HSS tree, as described in Section~\ref{sec:parcompr}. This is a serialized postorder traversal of the parallel part of the tree, where, at each node \(\tau\), we use the \texttt{PxGEMR2D} routine from ScaLAPACK to redistribute a block of the matrix with size \(n_\tau\times n_\tau\) from the \(p\) processes to the \(p_\tau\) processes that work at \(\tau\). The cost for one such redistribution is \([p,\frac{n_\tau^2}{p_\tau}]\) for the receiving processes and \([p_\tau,\frac{n_\tau^2}{p}]\) for the sending processes~\cite{prylli1997fast}. To get the total cost, we sum over the \(\mathcal{O}(p)\) nodes of the parallel part of the tree, and we use the fact that, at level \(i\) (0 being the root node), a node \(\tau\) is associated with two blocks of the original matrix with \(n_\tau=\frac{n}{2^i}\) rows and columns and is mapped on \(p_\tau=\frac{p}{2^i}\) processes. Each level has \(2^i\) nodes; at a given level, each process is receiver at one node (the node mapped on that process) and sender at \(2^i-1\) nodes. Therefore, the number of messages is
\[\sum_{\text{level }i=1}^{\log p} \left(1\cdot p+(2^i-1)\frac{p}{2^i}\right)
  =2p\log p-p\sum_{\text{level }i=1}^{\log p}\frac{1}{2^i}=p\log p-p\cdot\mathcal{O}(1)=\mathcal{O}(p\log p)\]
Similarly, the number of words to be transferred is:
\[\sum_{\text{level }i=1}^{\log p} \left(1\cdot \frac{(n/2^i)^2}{p/2^i}+(2^i-1)\frac{(n/2^i)^2}{p} \right) = \frac{n^2}{p}\sum_{\text{level }i=1}^{\log p}\frac{2^{i+1}-1}{2^{2i}}=\frac{n^2}{p}\cdot\mathcal{O}(1)=\mathcal{O}(\frac{n^2}{p})\]
Therefore, the cost for the initial distribution is, asymptotically, \([p\log p,\frac{n^2}{p}]\).
\item Postorder traversal of the tree to compute the generators. At a given node, there are three main ingredients:
\begin{enumerate}
\item Matrix-matrix products to compute the samples and updates. Using the above assumptions, all the blocks have size \(\mathcal{O}(r)\times \mathcal{O}(r\)) (e.g., \(2r\times r\)). The cost is thus \([r\log p_\tau,\frac{r^2}{\sqrt{p_\tau}}]\).
\item Interpolative decomposition of a block of size \(\mathcal{O}(r)\times\mathcal{O}(r\)) with rank \(\mathcal{O}(r)\); the cost is \([r\log p_\tau,r^2\frac{\log p_\tau}{\sqrt{p_\tau}}]\) (using Equation (4.1) from~\cite{wang2013efficient} with \(M=N=r\)).
\item Redistribution of blocks of size \(\mathcal{O}(r)\times\mathcal{O}(r\)) to the parent; the cost is \([1,\frac{r^2}{p_\tau}]\)~\cite{wang2013efficient}.
\end{enumerate}
The term corresponding to the redistribution (c) is negligible compared to the two other terms, and the term corresponding to Interpolative Decompositions (b) dominates the term corresponding to local matrix-matrix products (a). We sum (b) over the critical path (a branch of the tree). This time we number the levels so that the leaves of the parallel tree are at level 0, and the root is at level \(\log p\). At level \(i\), a node is mapped on \(p_i=2^i\) processes. The number of messages is
\[\sum_{i=1}^{\log p-1} r\log p_i=r\sum_{i=1}^{\log p-1}i=\mathcal{O}(r\log^2 p)\]
The number of words is, similarly,
\[\sum_{i=1}^{\log p-1} r^2\frac{\log p_i}{\sqrt{p_i}}=r^2\sum_{i=1}^{\log p-1}\frac{i}{2^{i/2}}=\mathcal{O}(r^2)\]
\end{enumerate}

We summarize the results in the following table:
\begin{table}[!ht]\centering
\renewcommand{\arraystretch}{1.8}
\begin{tabular}{|l|c|c|c|}\hline
Algorithm                      & Messages                           & Words \\
\hline\hline
ScaLAPACK \(LU\)               & \(\mathcal{O}(n\log p)\)                   & \(\mathcal{O}\left(n^2\frac{\log p}{\sqrt{p}}\right)\) \\
\hline
Non-randomized HSS compression & \(\mathcal{O}(p+r\log^2 p)\)               & \(\mathcal{O}\left(\frac{n^2}{p}+rn+r^2\log p\right)\) \\
\hline
\multirow{2}{*}{Randomized HSS compression}     & \(\mathcal{O}(p\log p+r\log p+r\log^2 p)\) & \(\mathcal{O}\left(\frac{n^2}{p}\ \ +\ \ \frac{rn}{\sqrt{p}}\ \ +\ \ r^2\right)\) \\
                               & \ \ dist\ \ \ \ GEMM\ \ \ \ \ tree & \ \ \ dist\ \ \ \ GEMM\ \ \ \ tree \\
\hline
\end{tabular}
\caption{Summary of communication costs.\label{tab:comm_costs}}
\end{table}

Now we take a closer look at the various communication costs in the randomized algorithm (last row of Table~\ref{tab:comm_costs}).
\begin{itemize}
\item In terms of latency, the initial distribution dominates for problems with small maximum rank, while the traversal of the tree dominates for problems with large rank.
\item In terms of bandwidth, when the rank is large, i.e., \(r>\mathcal{O}(\frac{n}{\sqrt{p}})\), the traversal of the tree dominates the matrix-matrix product, and the matrix-matrix product dominates the initial distribution. When \(r\) is small, i.e., \(r<\mathcal{O}(\frac{n}{\sqrt{p}})\), the initial distribution dominates the matrix-matrix product, and the matrix-matrix product dominates the traversal of the tree.
\end{itemize}

Comparing our randomized compression algorithm to ScaLAPACK LU, 
one can observe that, for problems with small rank \(r\), our algorithm communicates
fewer messages and less communication volume than ScaLAPACK does. However, for a large rank, 
it can be the opposite. We illustrate this in Section~\ref{sec:scal}.

Comparing our randomized compression algorithm to the non-randomized one previously 
developed, we observe the following:
\begin{itemize}
\item In terms of latency, our algorithm has a slightly larger complexity due to the \(\log p\) in the distribution term and the latency of the matrix-matrix product. We are investigating a way to reduce the number of messages to \(\mathcal{O}(p)\) in the initial distribution phase. For the matrix-matrix product, we could benefit from advances in communication-avoiding algorithms, such as the 2.5D matrix multiplication~\cite{solomonik2011communication}. 
\item In terms of bandwidth, for both compression algorithms, the first term
 corresponds to the initial distribution of the input matrix (Step (2) above). Afterwards, in the non-randomized HSS compression, there is a term for the \emph{row compression} (\(rn\)) and a term for the \emph{column compression} (\(r^2\log p\)). In the randomized algorithm, we have a term corresponding to the matrix-matrix product used for the sampling phase, and a term corresponding to the tree traversals. These terms are smaller than what appears in the communication cost of the non-randomized algorithm. Therefore, our randomized algorithm
 communicates fewer words, and we expect better performance in practice.
 We illustrate this in Section~\ref{sec:scal}.
\end{itemize}

\subsection{Parallel factorization, solution, and product}

The parallelization strategy for the factorization, triangular solution, and matrix-vector product are similar to the one we use for the compression. We exploit both tree parallelism, using a proportional mapping of the tasks, and node parallelism, by using PBLAS and ScaLAPACK operations. Serial subtrees are processed using sequential routines written using BLAS and LAPACK.

\section{Experimental results}
\label{sec:exps}

\subsection{Applications}
We report experimental results using the following matrices:
\begin{itemize}
\item \textbf{Toeplitz matrix:} a matrix \(A=\left[a_{i,j}\right]\) is a \emph{Toeplitz matrix} (or diagonal-constant matrix) if \(\forall (i,j), a_{i,j}=a_{i+1,j+1}\). We experimented with two Toeplitz matrices. The first one is a simple matrix with \(a_{i,i}=n^2\) and \(a_{i,j}=i-j\). It is diagonally dominant and yields very low HSS rank (a small constant). The second one is a kinetic energy matrix from quantum chemistry~\cite{jones2014development}; \(a_{i,i}=\frac{\pi^2}{6}\) and \(a_{i,j}=\frac{(-1)^{i-j}}{(i-j)^2d^2}\) where \(d\) is a discretization parameter (grid spacing). This matrix yields slightly larger maximum rank (that grows slowly with \(n\)) and is fairly ill-conditioned. This is a collaboration with D. J. Haxton and J. Jones at Lawrence Berkeley National Laboratory.
\item \textbf{Matrices from boundary element methods:} these matrices are known to be structured~\cite{hackbusch2000sparse,bebendorf2008hierarchical}. We obtained matrices from G. Sylvand (Airbus), and B. Notaros and A. Manic (Colorado State University). The matrices represent electromagnetic spheres (or collections of spheres). These matrices are known to be structured although the maximum rank is often large.
\item \textbf{Matrices from finite differences:} it is known that the \emph{inverse} of a sparse matrix arising from finite differences is dense and structured~\cite{hackbusch2000sparse,chandrasekaran2010numerical}. More specifically, the dense matrices that appear during sparse Gaussian Elimination are structured. Different approaches have been used to exploit this fact, especially in the context of the \emph{multifrontal method}~\cite{duff1983multifrontal}: using Block Low-Rank representations~\cite{amestoy2014improving}, Hierarchically Off-Diagonal Low-Rank matrices~\cite{aminfar2014fast} and HSS techniques~\cite{xia2013efficient,xia2013randomized,wang2014parallel}. In our experiments, we use dense matrices coming from the sparse factorization of the discretized Helmholtz equation; these matrices are generated by our code Hsolver~\cite{wang2014parallel}.
\item \textbf{Covariance matrices:} spatially correlated Gaussian random fields are useful in many modeling applications. They can be generated by solving an eigenvalue problem with a \emph{covariance matrix}~\cite{ladenheim2014multilevel}. These matrices are dense and are generally very large as they have as many degrees of freedom as the computational (physical) domain. However they are often compressible. We experimented with a covariance matrix generator provided by Panayot Vassilevski and Umberto Villa at the Lawrence Livermore National Lab, that relies on the MFEM code~\cite{kolevmfem}.
\item \textbf{\(\mathcal{H}-\)matrices:} we use an in-house matrix generator 
that produces \(\mathcal{H}-\)matrices (i.e., they have low-rank off-diagonal
 blocks but there is no recursive relation between the different blocks)
 with prescribed size. Such matrices can be compressed using HSS techniques;
 even though the maximum HSS rank will be larger than the maximum \(\mathcal{H}-\)rank,
 the compression can sometimes be done faster, depending on the problem.
\end{itemize}

We use two parallel machines at the National Energy Scientific Computing Center (NERSC).
 Hopper is a Cray XE6 system with 6384 nodes; each node has two twelve-core 2.1~GHz
 AMD Opteron 6172 processors and 32~GB of main memory.
 Edison is a Cray XC30 system; each node has two twelve-core 2.4~GHz 
Intel Xeon E5-2695 processors and 64~GB of main memory.

\subsection{General trees}
\label{sec:comb}

In most of our examples and experiments, we use complete binary trees. Here, we briefly illustrate that our code can handle more general trees. This is important, as, in some applications, the clustering of the variables might not be a straightforward recursive bisection, thus the tree might not be balanced. Here, we use an example where the tree is a binary ``comb'', i.e.,  for each pair siblings, only one of the siblings has children. Hilbert matrices exhibit such a structure~\cite{bebendorf2008hierarchical}.

The matrix we use is an \(\mathcal{H}-\)matrix with size \(40,000\times40,000\). It has the structure illustrated in Figure~\ref{fig:combmat}, corresponding to the comb tree in Figure~\ref{fig:combtree}.

\begin{figure}[!ht]
\centering
\subfigure[Matrix structure.\label{fig:combmat}]{
\begin{tikzpicture}[scale=0.075]
\draw[thick] (1,-1) -- ++(-2,0) -- ++(0,49) -- ++(2,0);
\draw[thick] (46,-1) -- ++(2,0) -- ++(0,49) -- ++(-2,0);
\fill (0,42) rectangle +(5,5);
\fill (6,36) rectangle +(5,5);
\fill (12,24) rectangle +(11,11);
\fill (24,0) rectangle +(23,23);
\fill[gray!60] (0,36) rectangle +(5,5);
\fill[gray!60] (6,42) rectangle +(5,5);
\fill[gray!60] (0,24) rectangle +(11,11);
\fill[gray!60] (12,36) rectangle +(11,11);
\fill[gray!60] (0,0) rectangle +(23,23);
\fill[gray!60] (24,24) rectangle +(23,23);
\end{tikzpicture}
}
\subfigure[Ranks.\label{fig:combtree}]{
\begin{tikzpicture}[scale=0.79]
\tikzstyle{lnode} = [draw, circle,fill=black,scale=1.2]
\tikzstyle{inode} = [draw, circle,fill=gray!60,scale=1.2]
\tikzstyle{level 1} = [label distance=-3.5,level distance=0.08\textwidth, sibling distance=0.09\textwidth]
\tikzstyle{level 3} = [label distance=-0]
  \node[inode](7){}
  child{node[inode,label=north west:400](5){}
    child{node[inode,label=north west:600](3){}
      child{node[lnode,label=south:700](1){}}
      child{node[lnode,label=south:700](2){}}
    }
    child{node[lnode,label=north east:600](4){}
    }
  }
  child{node[lnode,label=north east:400](6){}
  };
\end{tikzpicture}
}
\subfigure[Uniform mapping.\label{fig:combmap}]{
\begin{tikzpicture}[scale=0.79]\small
\tikzstyle{lnode} = [draw, circle,fill=black,scale=1.2]
\tikzstyle{inode} = [draw, circle,fill=gray!60,scale=1.2]
\tikzstyle{level 1} = [label distance=-3.5,level distance=0.08\textwidth, sibling distance=0.09\textwidth]
\tikzstyle{level 3} = [label distance=-0]
  \node[inode,label=north:64](7){}
  child{node[inode,label=north west:32](5){}
    child{node[inode,label=north west:16](3){}
      child{node[lnode,label=south:8](1){}}
      child{node[lnode,label=south:8](2){}}
    }
    child{node[lnode,label=north east:16](4){}
    }
  }
  child{node[lnode,label=north east:32](6){}
  };
\end{tikzpicture}
}
\subfigure[Weighted mapping.\label{fig:combremap}]{
\begin{tikzpicture}[scale=0.79]\small
\tikzstyle{lnode} = [draw, circle,fill=black,scale=1.2]
\tikzstyle{inode} = [draw, circle,fill=gray!60,scale=1.2]
\tikzstyle{level 1} = [label distance=-3.5,level distance=0.08\textwidth, sibling distance=0.09\textwidth]
\tikzstyle{level 3} = [label distance=-0]
  \node[inode,label=north:64](7){}
  child{node[inode,label=north west:16](5){}
    child{node[inode,label=north west:4](3){}
      child{node[lnode,label=south:2](1){}}
      child{node[lnode,label=south:2](2){}}
    }
    child{node[lnode,label=north east:12](4){}
    }
  }
  child{node[lnode,label=north east:48](6){}
  };
\end{tikzpicture}
}
\caption{Structured matrix with a comb-shaped clustering tree~\subref{fig:combmat}. HSS compression with a comb-shaped HSS tree yields low maximum rank~\subref{fig:combtree}. The tree can be mapped to processes using a uniform mapping where pairs of siblings are mapped on the same number of processes~\subref{fig:combmap} or a mapping that assigns more processes to right nodes~\subref{fig:combremap}. For example, in~\subref{fig:combmap}, the children of the root node are both mapped on 32 processes, but in~\subref{fig:combremap} and they are mapped on 16 and 48 processes.}
\label{fig:comb}
\end{figure}

\begin{figure}[!ht]
\centering
\subfigure[Matrix structure.\label{fig:compmat}]{
\begin{tikzpicture}[scale=0.075]\small
\draw[thick] (1,-1) -- ++(-2,0) -- ++(0,49) -- ++(2,0);
\draw[thick] (46,-1) -- ++(2,0) -- ++(0,49) -- ++(-2,0);
\fill (0,42) rectangle +(5,5);
\fill (6,36) rectangle +(5,5);
\fill (12,24) rectangle +(5,5);
\fill (18,24) rectangle +(5,5);
\fill (12,30) rectangle +(5,5);
\fill (18,30) rectangle +(5,5);
\foreach \i in {24,30,36,42}{
\foreach \j in {0,6,12,18}{
\fill (\i,\j) rectangle +(5,5);
}
}
\fill[gray!60] (0,36) rectangle +(5,5);
\fill[gray!60] (6,42) rectangle +(5,5);
\fill[gray!60] (0,24) rectangle +(11,11);
\fill[gray!60] (12,36) rectangle +(11,11);
\fill[gray!60] (0,0) rectangle +(23,23);
\fill[gray!60] (24,24) rectangle +(23,23);
\fill[pattern color=red,pattern=north west lines](0,12) rectangle +(23,11);
\fill[pattern color=red,pattern=north west lines](36,12) rectangle +(11,11);
\fill[pattern color=red,pattern=north west lines](24,24) rectangle +(11,23);
\fill[pattern color=red,pattern=north west lines](24,0) rectangle +(11,11);
\hfill
\end{tikzpicture}
}
\subfigure[Tree structure and ranks.\label{fig:comptree}]{
\begin{tikzpicture}[scale=0.75]\small
\tikzstyle{lnode} = [draw, circle,fill=black,scale=1.2]
\tikzstyle{inode} = [draw, circle,fill=gray!60,scale=1.2]
\tikzstyle{level 1} = [label distance=-3.5,level distance=0.08\textwidth, sibling distance=0.36\textwidth]
\tikzstyle{level 2} = [label distance=-3.5,level distance=0.08\textwidth, sibling distance=0.18\textwidth]
\tikzstyle{level 3} = [label distance=0,level distance=0.08\textwidth, sibling distance=0.09\textwidth]
  \node[lnode](15){}
  child{node[inode,label=north west:400](7){}
    child{node[inode,label=north west:600](3){}
      child{node[lnode,label=south:700](1){}}
      child{node[lnode,label=south:700](2){}}
    }
    child{node[lnode,label=north east:600](6){}
      child{node[lnode,label=south:5000](4){}}
      child{node[lnode,label=south:5000](5){}}
    }
  }
  child{node[lnode,label=north east:400](14){}
    child{node[preaction={fill=black},lnode,pattern=north west lines,pattern color=red,label=north west:10000](10){}
      child{node[lnode,label=south:5000](8){}}
      child{node[lnode,label=south:5000](9){}}
    }
    child{node[lnode,label=north east:10000](13){}
      child{node[lnode,label=south:5000](11){}}
      child{node[lnode,label=south:5000](12){}}
    }
  };
\end{tikzpicture}
}
\caption{Matrix from Figure \ref{fig:comb} compressed using a complete binary tree.}
\label{fig:comp}
\end{figure}
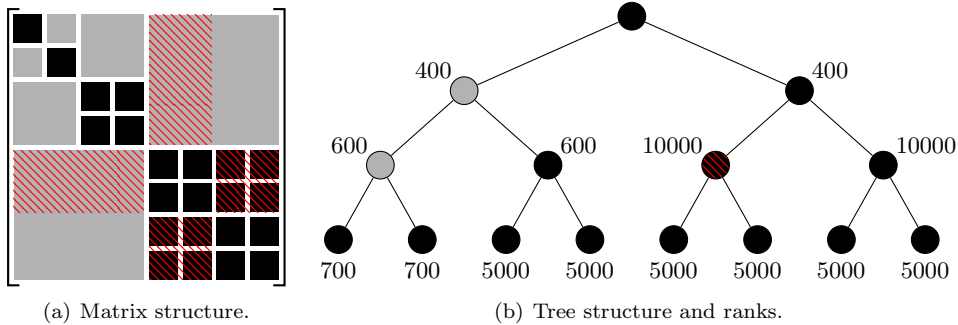

In Table~\ref{tab:comb}, we report some experiments with this matrix. We compare the effect of using a comb-shaped tree instead of a binary tree for the HSS compression, and we illustrate that we can modify the weights used in the proportional mapping to improve performance. In the first experiment, the HSS compression is based on a binary tree with 4 levels shown in Figure~\ref{fig:comptree}. One can easily understand why the maximum rank is \(\frac{40000}{4}=10000\); it comes from the fact that the (2,2) block of the matrix, of size \(20,000\times20,000\) is not structured, i.e., not HSS compressible. Its off-diagonal blocks, of size \(10,000\times10,000\) are full-rank. This yields the maximum rank because, at the striped node in Figure~\ref{fig:comptree}, the blocks to be compressed are the striped blocks in Figure~\ref{fig:compmat} and they have rank 10,000 because they contain full-rank blocks (black in the figure).

In the second experiment, the HSS compression is based on a comb tree with the same structure as in Figure~\ref{fig:combtree}. The compression is much faster and the maximum rank is only 700 (a parameter of our example).

Finally, the last experiment consists in remapping the HSS tree by modifying the weights used in the proportional mapping. Instead of using even weights (which yield an even splitting of processes between the left and right branches of the tree), we choose to attribute more processes to right nodes. Whenever a pair of siblings is mapped, the right child inherits from 75\% of the processes working at its parent. This is motivated by the fact that, in the factorization, we have to perform the \(LQ\) factorization of a \(20,000\times20,000\) block (corresponding to the (2,2) block of the original matrix), which is the most costly operation in the factorization and corresponds to the right child of the root node. By simply changing the weights in the mapping procedure using some knowledge of the input matrix, we significantly speed-up the factorization (65.8 seconds instead of 101.4 seconds), at the price of a small increase in the compression time (19.5 seconds instead of 14.1).

\begin{table}[!ht]
\centering
\begin{tabular}{|l|r|r|r|}
\hline
                           & Binary tree & Comb tree & Comb tree, remapped \\
\hline
HSS compression time (s)   &       704.1 &      14.1 &                19.5 \\
Maximum rank               &       10000 &       700 &                 700 \\
ULV factorization time (s) &       122.0 &     101.4 &                65.8 \\
\hline
\end{tabular}
\caption{Experiments with the structured matrix in Figure~\ref{fig:comb} using 64 MPI tasks.~\label{tab:comb}}
\end{table}

This simple example illustrates that our code is flexible. It can handle different tree structures, and different task-to-process mappings, using any number of processes. In Hsolver, this was not the case. The code was restricted to some specific problems, relied on complete binary trees, and could only work with power-of-two number of processes.

\subsection{Solving linear systems}

In this section, we illustrate the performance of our code for seven different matrices from the abovementioned applications. The results are reported in Table~\ref{tab:resSolve}. For each matrix, we experiment with four different compression thresholds (\(\varepsilon=10^{-8},10^{-6},10^{-4},10^{-2}\)) and we report statistics for the HSS compression, the ULV factorization and the triangular solution with iterative refinement. We provide run time, number of floating-point operations, size of the HSS/ULV factors, and we compare with the run time for solving a system with ScaLAPACK. In terms of memory, ignoring small temporary storage and communication buffers, the memory footprint for ScaLAPACK is simply the storage of the matrix \(A\). In our new code, the memory usage consists not only of the original matrix, but also storage for the random vectors and the samples, and storage for the HSS and ULV factors. We report a \emph{memory overhead}, which represents the extra memory usage of STRUMPACK relative to that of ScaLAPACK.

Among this collection of matrices, the data type for the two matrices \textsc{BEMMultiSphere} and \textsc{Schur100} is single precision. The other matrices are of double precision. For the single precision input, the compression threshold \(10^{-8}\) is very small, leading to almost no compression. For example, for matrix \textsc{Schur100}, the (1,2) block is of size 5,000, whereas the maximum rank is 4933 with \(\varepsilon = 10^{-8}\), which is essentially full rank. The memory overhead is much larger with HSS representation. This is mostly due to the ULV factors. Indeed, the special structure of the \(U\) and \(V\) generators keep memory usage low when blocks are full-rank, and the HSS form has roughly the same memory footprint as the input matrix. However, in this situation, ULV factors are usually much larger than the HSS form. Therefore, the practical use of HSS algorithms is not with very small compression threshold \(\varepsilon\).

\begin{table}[!ht]
\setlength{\tabcolsep}{0.18em}
\renewcommand{\arraystretch}{1.25}
\newcommand{\ce}[1]{\multicolumn{1}{c|}{#1}}
\newcommand{\cee}[1]{\multicolumn{1}{c||}{#1}}
\newcommand{\mr}[1]{\multirow{4}{*}{#1}}
\newcommand{\mrr}[1]{\multirow{2}{*}{#1}}
\resizebox{\textwidth}{!}{
\begin{tabular}{|l|c|r||r|r|r|r||r|r|r||r|r||r|r||r|r|}
\hhline{---||---------||--||--}
Matrix                       &     Size    & \cee{\(\varepsilon\)} & \multicolumn{9}{c||}{STRUMPACK: solution with HSS compression and factorization} & \multicolumn{2}{c||}{Solution with} & \multicolumn{2}{c|}{Comparison: HSS} \\
\hhline{|~|~|~||---------||~~~|}
                             &             &             & \multicolumn{4}{c||}{HSS compression} & \multicolumn{3}{c||}{ULV factorization} & \multicolumn{2}{c||}{Solution+IR} & \multicolumn{2}{c||}{ScaLAPACK} & \multicolumn{2}{c|}{vs ScaLAPACK}\\
\hhline{|~|~|~||-|-|-|-||-|-|-||-|-||-|-||-|-|}
                             &             &             &  Max &  Factors  &    \ce{Flops}    &   Time    &  Factors  &    \ce{Flops}    &   Time    &    \ce{Flops}    &   Time    &       \ce{Flops}    &     Time   &   Memory &     \ce{Time} \\
                             &             &             & rank & \ce{(MB)} & (\(\times10^9\)) & \cee{(s)} & \ce{(MB)} & (\(\times10^9\)) & \cee{(s)} & (\(\times10^9\)) & \cee{(s)} & (\(\times10^{12}\)) &  \cee{(s)} & overhead &       speedup \\
\hhline{===::====::===::==::==::==}
                             & \mr{80,000} & \(10^{-8}\) &    2 &      14.6 &            307.3 &      11.4 &      37.2 &             0.10 &     0.008 &            0.008 &       0.2 &          \mr{341.3} & \mr{856.6} &    0.1\% & \textbf{75.2} \\
\hhline{~~-||----||---||--||~~--}
\textsc{Simple}              &             & \(10^{-6}\) &    2 &      14.2 &            307.3 &      11.3 &      37.0 &             0.10 &     0.007 &            0.025 &       1.2 &                     &            &    0.1\% &          68.1 \\
\hhline{~~-||----||---||--||~~--}
\textsc{Toeplitz}            &             & \(10^{-4}\) &    3 &      13.3 &            307.3 &      11.3 &      36.3 &             0.09 &     0.006 &            0.049 &       2.8 &                     &            &    0.1\% &          60.5 \\
\hhline{~~-||----||---||--||~~--}
                             &             & \(10^{-2}\) &    3 &      13.2 &            307.3 &      11.3 &      36.2 &             0.09 &     0.006 &            0.088 &       5.4 &                     &            &    0.1\% &          51.0 \\
\hhline{===::====::===::==::==::==}
                             & \mr{80,000} & \(10^{-8}\) &  169 &      55.1 &           6411.2 &      19.0 &     152.7 &             1.40 &      0.04 &              1.0 &       1.5 &          \mr{341.3} & \mr{894.1} &    1.5\% &          43.5 \\
\hhline{~~-||----||---||--||~~--}                                                                                                                                                                                                 
\textsc{QChem}               &             & \(10^{-6}\) &  147 &      42.1 &           5126.6 &      17.6 &     110.0 &             0.86 &      0.03 &              0.9 &       2.0 &                     &            &    1.4\% & \textbf{45.5} \\
\hhline{~~-||----||---||--||~~--}                                                                                                                                                                                                 
\textsc{Toeplitz}            &             & \(10^{-4}\) &  120 &      33.3 &           3843.7 &      16.7 &      83.6 &             0.58 &      0.02 &              1.7 &       5.6 &                     &            &    1.0\% &          40.0 \\
\hhline{~~-||----||---||--||~~--}                                                                                                                                                                                                 
                             &             & \(10^{-2}\) &   30 &      18.1 &           1280.6 &      13.4 &      42.7 &             0.12 &      0.01 &              N/A &       N/A &                     &            &    0.5\% &           N/A \\
\hhline{===::====::===::==::==::==}
\mr{\textsc{HMatrix}}        & \mr{80,000} & \(10^{-8}\) &  787 &    2235.3 &          24707.5 &      52.5 &    7897.7 &           1865.0 &       3.5 &              2.1 &      0.22 &          \mr{341.3} & \mr{862.1} &   10.3\% &          15.3 \\
\hhline{~~-||----||---||--||~~--}
                             &             & \(10^{-6}\) &  785 &    2263.4 &          24723.3 &      53.8 &    7966.9 &           1897.9 &       3.5 &              4.3 &      0.77 &                     &            &   10.3\% &          14.9 \\
\hhline{~~-||----||---||--||~~--}
                             &             & \(10^{-4}\) &    4 &      14.3 &            409.7 &      11.1 &      37.1 &              0.1 &     0.006 &             0.02 &      0.72 &                     &            &    0.2\% & \textbf{72.9} \\
\hhline{~~-||----||---||--||~~--}
                             &             & \(10^{-2}\) &    2 &      13.2 &            409.7 &      11.1 &      36.2 &              0.1 &     0.006 &             0.02 &      1.26 &                     &            &    0.2\% &          69.7 \\
\hhline{===::====::===::==::==::==}
                             & \mr{10,002} & \(10^{-8}\) & 1433 &     722.0 &            972.4 &       9.6 &    2029.2 &            401.8 &       3.1 &              0.5 &      0.07 &            \mr{0.7} &   \mr{3.3} &  120.1\% &           0.3 \\
\hhline{~~-||----||---||--||~~--}
\textsc{BEM}                 &             & \(10^{-6}\) & 1016 &     507.6 &            624.3 &       4.7 &    1265.5 &            172.6 &       1.7 &              0.5 &       0.1 &                     &            &   82.4\% &           0.5 \\
\hhline{~~-||----||---||--||~~--}
\textsc{Acoustics}           &             & \(10^{-4}\) &  793 &     420.0 &            453.6 &       3.3 &     988.8 &            109.2 &       1.1 &              0.7 &       0.3 &                     &            &   66.6\% &  \textbf{0.7} \\
\hhline{~~-||----||---||--||~~--}
                             &             & \(10^{-2}\) &  379 &     288.4 &            243.1 &       1.7 &     667.4 &             54.4 &       0.5 &              N/A &       N/A &                     &            &   44.9\% &           N/A \\
\hhline{===::====::===::==::==::==}
\mrr{\textsc{BEM}}           & 27,648 & \(10^{-8}\) & 5995 &    4489.0 &          38980.1 &     354.5 &   15897.3 &          26406.2 &      60.4 &              8.9 &       0.9 &           \mr{14.1} &  \mr{30.7} &  403.2\% &          0.07 \\
\hhline{~~-||----||---||--||~~--}
\mrr{\textsc{Multi}}         &\mrr{(Single}    & \(10^{-6}\) & 2145 &     811.7 &           8499.1 &      29.1 &    2568.9 &           1015.6 &       3.4 &              1.5 &       0.5 &                     &            &   53.5\% &           0.9 \\
\hhline{~~-||----||---||--||~~--}
\mrr{\textsc{Sphere}}        & \mrr{prec.)}   & \(10^{-4}\) & 1488 &     425.6 &           5396.0 &      15.0 &    1237.9 &            322.8 &       1.2 &              1.0 &       0.7 &                     &            &   38.4\% &           1.8 \\
\hhline{~~-||----||---||--||~~--}
                             &             & \(10^{-2}\) &  800 &     184.0 &           3179.7 &       8.5 &     495.3 &             52.2 &       0.3 &              1.0 &       1.8 &                     &            &   25.2\% &  \textbf{2.9} \\
\hhline{===::====::===::==::==::==}
\mr{\textsc{Schur100}}       & {10,000} & \(10^{-8}\) & 4933 &     763.0 &           4520.4 &      98.3 &    2957.1 &           2928.8 &      19.8 &             16.3 &       3.6 &            \mr{0.7} &   \mr{4.2} &  722.6\% &          0.03 \\
\hhline{~~-||----||---||--||~~--}
                             & \mrr{(Single}      & \(10^{-6}\) &  840 &     136.2 &            601.0 &       6.3 &     388.3 &             61.0 &       0.9 &              2.2 &       1.7 &                     &            &   76.6\% &           0.5 \\
\hhline{~~-||----||---||--||~~--}
                             & \mrr{prec.)}  & \(10^{-4}\) &  501 &      90.3 &            278.3 &       3.2 &     235.2 &             23.0 &       0.5 &              1.3 &       1.4 &                     &            &   40.5\% &           0.8 \\
\hhline{~~-||----||---||--||~~--}
                             &             & \(10^{-2}\) &  282 &      56.6 &            153.0 &       2.0 &     134.0 &              7.4 &       0.2 &              1.3 &       2.5 &                     &            &   25.6\% &  \textbf{0.9} \\
\hhline{===::====::===::==::==::==}
\mr{\textsc{Covar30}}        & \mr{27,000} & \(10^{-8}\) & 2247 &    1221.6 &           8976.3 &      34.3 &    3157.2 &           1515.2 &       4.6 &              0.4 &       0.1 &           \mr{13.1} &  \mr{36.8} &   61.1\% &           0.9 \\
\hhline{~~-||----||---||--||~~--}
                             &             & \(10^{-6}\) & 1609 &     948.2 &           6363.7 &      18.6 &    2301.7 &            815.8 &       2.7 &              3.2 &       0.8 &                     &            &   47.7\% &  \textbf{1.7} \\
\hhline{~~-||----||---||--||~~--}
                             &             & \(10^{-4}\) &  215 &     380.8 &           1093.2 &       3.4 &    1054.7 &            212.6 &       0.6 &             N/A  &       N/A &                     &            &   14.3\% &           N/A \\
\hhline{~~-||----||---||--||~~--}
                             &             & \(10^{-2}\) &    3 &     348.0 &             49.6 &       1.9 &    1042.8 &            205.4 &       0.5 &             N/A  &       N/A &                     &            &    6.7\% &           N/A \\
\hhline{---||----||---||--||--||--}
\end{tabular}
}
\caption{Solving linear systems from different applications using 64 MPI tasks.~\label{tab:resSolve}}
\end{table}

All the problems exhibit the same -- and expected -- behavior. When the compression threshold \(\varepsilon\) is higher (e.g., \(10^{-2}\)), the compression and the factorization are faster and the HSS and ULV factors are smaller than when \(\varepsilon\) is closer to machine precision. The gains in compression and factorization come at the price of accuracy; for some problems, the solution is inaccurate when \(\varepsilon\) is too large. This is the case for matrices \textsc{QChemToeplitz}, \textsc{BEMAcoustics} and \textsc{Covar30}. For some other problems, accuracy is satisfying with the largest value of \(\varepsilon\), but the best choice of \(\varepsilon\) is not \(10^{-2}\). For example, for problem \textsc{Hmatrix}, the best choice is \(\varepsilon=10^{-4}\).

We now compare the behavior of our dense solver with ScaLAPACK. The last column in Table~\ref{tab:resSolve} is the speed-up of STRUMPACK with respect to ScaLAPACK. For synthetic problems (\textsc{SimpleToeplitz}, \textsc{HMatrix}) and problems with a very simple structure (\textsc{QChemToeplitz}), using HSS techniques yields very large gains. For example, for the \textsc{HMatrix} problem and \(\varepsilon=10^{-4}\), our solution process is 72.9 times faster than a traditional dense \(LU\) factorization. For problems \textsc{BEMMultiSphere} and \textsc{Covar30}, the gains are less impressive but still significant; STRUMPACK exhibits a 2.9x speed-up for \textsc{BEMMultiSphere} and a 1.7x speed-up in run time for \textsc{Covar30}. For the last two problems, \textsc{Schur100} and \textsc{BEMAcoustics}, STRUMPACK is slower than ScaLAPACK regardless of the parameters. These two matrices exhibit some low-rank property and the number of floating-point operations performed by STRUMPACK is lower than that of ScaLAPACK, but, however, the total run time is larger with STRUMPACK. This highlights a drawback of our approach. Traditional dense \(LU\) factorization is an algorithm with a very regular computational pattern, than can be written with BLAS3 kernels, and good implementations (e.g., vendor-tuned) usually exhibit very good flop-rate and can reach a very large fraction of the peak performance. On the other hand, algorithms that takes advantage of low-rank structures (e.g., HSS, but also \(\mathcal{H}\)-matrices or Block Low-Rank representations) have to deal with more irregular and imbalanced task flows, and manipulate a collection of small matrices instead of one large matrix. Therefore, these algorithms cannot be expected to reach the same flop rate as traditional algorithms. This is visible in Table~\ref{tab:resSolve}. 

We want to highlight that for a given class of applications, using low-rank approximation techniques usually pay off past a certain size. This is because, 
although HSS techniques allow to solve linear systems with a lower 
asymptotic complexity, but with a larger constant prefactor. Also, as we mentioned previously, the flop rate with HSS is often lower than with traditional algorithms. These effects are visible in Section~\ref{sec:scal} where we experiment with matrices of growing size from a particular application; in this framework, gains increase with problem size.

The last point that we elaborate on is memory. As stated previously, the \emph{memory overhead} is the extra memory usage of STRUMPACK relative to that of ScaLAPACK; calling \(mem_{sca}\) the memory usage of ScaLAPACK and \(mem_{str}\) the memory usage of STRUMPACK, this is simply \(\frac{mem_{str}-mem_{sca}}{mem_{str}}\). It is important to understand that this memory overhead also represents the amount of memory that would be used if we were to use a matrix-free implementation. We recall that our algorithm is amenable to a matrix-free framework since it only requires access to a matrix-vector routine and \emph{selected} elements, more specifically \(\mathcal{O}(r^2 n)\), elements of the matrix (with \(r\) the maximum rank and assuming the tree has \(\log n\) levels). For example, for matrix \textsc{BEMMultiSphere}, the memory overhead is 25.2\%, which means:
\begin{enumerate}
\item The memory consumption of STRUMPACK is 1.252 times that of ScaLAPACK. 
\item If we were to use a matrix-free version the memory consumption would be
 25.2\% that of ScaLAPACK, i.e., a 4-fold reduction.
\end{enumerate}

\subsection{Fast matrix-vector product}

In this section, we briefly illustrate the use of HSS techniques for fast matrix-vector products. Here, the matrix is not factored with ULV but is simply kept in HSS form to perform matrix-vector multiplication. We use the power method (that relies mainly on matrix-vector products) to compute the largest eigenvalue of the \textsc{QChemToeplitz} matrix.

\begin{table}[!ht]
\setlength{\tabcolsep}{0.25em}
\begin{tabular}{|r|r|r|r|r|r|r||r|r|r||r|}
\hhline{|-|-|-|-|-|-|-||-|-|-||-|}
\multicolumn{7}{|c||}{HSS} & \multicolumn{3}{c||}{Traditional GEMV} & \multicolumn{1}{c|}{Speed-up} \\
\hhline{|-|-|-|-|-|-|-||-|-|-||~|}
\multicolumn{4}{|c|}{Compression} & \multicolumn{3}{c||}{Iterations} & \multicolumn{3}{c||}{Iterations} & \multicolumn{1}{c|}{with HSS} \\
\hhline{|-|-|-|-|-|-|-||-|-|-||-|}
Max rank & Factors (MB) & Flops (\(\times10^9\)) & Time & \#It & Flops (\(\times10^9\)) & Time & \#It &  Flops & Time &  \\
\hhline{|-|-|-|-|-|-|-||-|-|-||-|}
147 & 42.1 & 5126.6 & 17.6 & 318 & 1053.0 & 21.8 & 318 & 4070.4 & 69.7 & 1.8 \\
\hhline{|-|-|-|-|-|-|-||-|-|-||-|}
\end{tabular}
\caption{Power method for QChemToeplitz using 64 MPI tasks.~\label{tab:resPower}}
\end{table}

\subsection{Adaptive-rank mechanism}

In this section, we illustrate the behavior of the adaptive sampling mechanism described in Section~\ref{sec:adaptive}. The matrix we use corresponds to an electromagnetic sphere discretized with the boundary element method and has size 58,800. 

\begin{figure}[!ht]\centering
\subfigure[In steps of 500.\label{fig:adapt500}]{\resizebox{0.375\textwidth}{!}{
\renewcommand{\arraystretch}{0.9}
\begin{tikzpicture}[scale=0.75]\small
\tikzstyle{cnode} = [draw, circle,scale=1.2]
\tikzstyle{level 1} = [label distance=-3.5,level distance=0.09\textwidth, sibling distance=0.16\textwidth]
\tikzstyle{level 2} = [label distance=0,level distance=0.09\textwidth, sibling distance=0.08\textwidth]
  \node[cnode](7){}
    child{node[cnode,label={[yshift=0.35cm]west:\begin{tabular}{r}-\\-\\-\\-\\-\\\textbf{2306}\end{tabular}}](3){}
      child{node[cnode,label=south:\begin{tabular}{r}\underline{500}\\\underline{1000}\\\underline{1435}\\\textbf{1592}\\-\end{tabular}](1){}}
      child{node[cnode,label=south:\begin{tabular}{r}\underline{500}\\\underline{1000}\\\underline{1500}\\\underline{1982}\\\underline{2358}\\\textbf{2479}\end{tabular}](2){}}
    }
    child{node[cnode,label={[yshift=0.35cm]east:\begin{tabular}{r}-\\-\\-\\-\\-\\\textbf{2386}\end{tabular}}](6){}
      child{node[cnode,label=south:\begin{tabular}{r}\underline{500}\\\underline{1000}\\\underline{1500}\\\underline{1993}\\\underline{2401}\\\textbf{2673}\end{tabular}](4){}}
      child{node[cnode,label=south:\begin{tabular}{r}\underline{500}\\\underline{1000}\\\underline{1452}\\\textbf{1660}\\-\end{tabular}](5){}}
    }
  ;
\draw (3)++(-2.2,0.66) node[inner sep=0,outer sep=0]{\begin{tabular}{r}Rand.\\500\\1000\\1500\\2000\\2500\\3000\end{tabular}};
\draw (3)++(-2.2,-2.7) node[inner sep=0,outer sep=0]{\begin{tabular}{r}Rand.\\500\\1000\\1500\\2000\\2500\\3000\end{tabular}};
\end{tikzpicture}
}}%
\subfigure[In steps of 1,000.\label{fig:adapt1000}]{\resizebox{0.375\textwidth}{!}{
\renewcommand{\arraystretch}{0.9}
\begin{tikzpicture}[scale=0.75]\small
\tikzstyle{cnode} = [draw, circle,scale=1.2]
\tikzstyle{level 1} = [label distance=-3.5,level distance=0.09\textwidth, sibling distance=0.16\textwidth]
\tikzstyle{level 2} = [label distance=0,level distance=0.09\textwidth, sibling distance=0.08\textwidth]
  \node[cnode](7){}
    child{node[cnode,label=west:\begin{tabular}{r}-\\-\\\textbf{2285}\end{tabular}](3){}
      child{node[cnode,label=south:\begin{tabular}{r}\underline{1000}\\\textbf{1603}\\-\end{tabular}](1){}}
      child{node[cnode,label=south:\begin{tabular}{r}\underline{1000}\\\underline{1982}\\\textbf{2560}\end{tabular}](2){}}
    }
    child{node[cnode,label=east:\begin{tabular}{r}-\\-\\\textbf{2336}\end{tabular}](6){}
      child{node[cnode,label=south:\begin{tabular}{r}\underline{1000}\\\underline{1992}\\\textbf{2648}\end{tabular}](4){}}
      child{node[cnode,label=south:\begin{tabular}{r}\underline{1000}\\\textbf{1610}\\-\end{tabular}](5){}}
    }
  ;
\draw (3)++(-2.2,0.21) node[inner sep=0,outer sep=0]{\begin{tabular}{r}Rand.\\1000\\2000\\3000\end{tabular}};
\draw (3)++(-2.2,-2.09) node[inner sep=0,outer sep=0]{\begin{tabular}{r}Rand.\\1000\\2000\\3000\end{tabular}};
\end{tikzpicture}
}}%
\subfigure[No adaptive sampling.\label{fig:noadapt}]{\resizebox{0.24\textwidth}{!}{
\begin{tikzpicture}[scale=0.75]\small
\tikzstyle{cnode} = [draw, circle,scale=1.2]
\tikzstyle{level 1} = [label distance=-3.5,level distance=0.09\textwidth, sibling distance=0.16\textwidth]
\tikzstyle{level 2} = [label distance=0,level distance=0.09\textwidth, sibling distance=0.08\textwidth]
  \node[cnode](7){}
    child{node[cnode,label=north west:2281](3){}
      child{node[cnode,label=south:1668](1){}}
      child{node[cnode,label=south:2569](2){}}
    }
    child{node[cnode,label=north east:2334](6){}
      child{node[cnode,label=south:2661](4){}}
      child{node[cnode,label=south:1730](5){}}
    }
  ;
\end{tikzpicture}
}
}
\caption{Ranks of the \(U\) generators using adaptive sampling in steps of 500~\subref{fig:adapt500}, in steps of 1,000~\subref{fig:adapt1000}, and without adaptive sampling~\subref{fig:noadapt}. An underlined rank means that the corresponding node triggers a restart. A rank marked in bold is final.}
\label{fig:adaptive}
\end{figure}
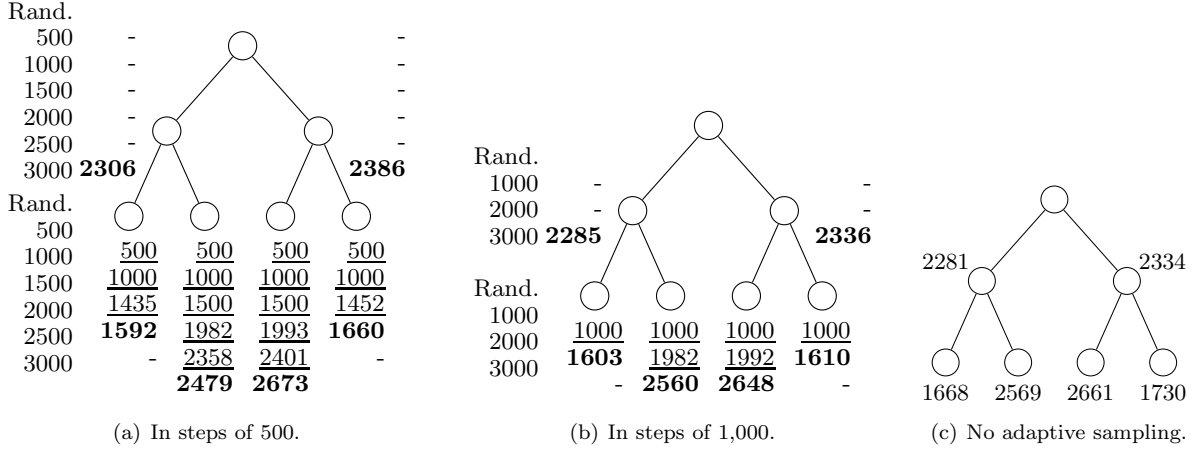

In Figure~\ref{fig:adaptive}, we examine three configurations. In Figure~\ref{fig:noadapt}, we use 3,000 random vectors, which is enough to guarantee that we reveal the ``true'' rank of each node (in this discussion we only consider the rank of the \(U\) generator, for simplicity). In Figure~\ref{fig:adapt500}, we start the compression with 500 random vectors. Every time a block is compressed, we look at the difference between its rank and the number of random vectors; if it was less than 200, we discarded the generators at the node we consider, we add 500 new random vectors, and the compression restarts. In this example, the four leaves of the tree are compressed in parallel; they all have rank 500, and the compression restarts with \(500+500=1,000\) random vectors. Again, this is not enough (rank 1000 is found at the leaves), and we add 500 new random vectors. 1,000 random vectors is not enough and the compression restarts with \(1,000+500=1,500\) random vectors; this is is enough, and the compression restarts with \(1500+500=2000\) random vectors. This time, the leaves exhibit four different ranks: 1,592, 1,982, 1,993, and 1,660. At the leaves with 1,592 and 1,660, the generators are kept because the difference between their rank and the number of random vectors is more than the limit that we picked. However, the compression needs to restart because of the two other leaves. We add 500 random vectors and the traversal restarts. At the leaves that have rank 1,592 and 1,660, we simply update the samples \(S^r_\tau\) and \(S^c_\tau\); their generators have been obtained at the previous iteration and are not recomputed. At the two other leaves, we recompute the generators. This fails again and the compression restarts with 3,000 random vectors. This time the ranks are small enough and the generators are kept. The compression proceeds to the next level then terminates.

In Figure~\ref{fig:adapt1000}, we start with 1,000 random vectors and we add 1,000 random vectors whenever a step of compression fails; this time, the compression restarts only twice and successfully terminates with 3,000 random vectors. One can observe that the ranks that we obtain using different numbers of random vectors vary slightly. This is an effect of the sampling mechanism, but it does not have any major effect on accuracy, or the size of the HSS and ULV representations. However, the adaptive sampling mechanism influences performance. In Table~\ref{tab:adaptive}, we report on the run time for the HSS compression when the tree has 3 levels (as in Figure~\ref{fig:adaptive}) and when the tree has 8 levels. One can observe that when the HSS tree has 3 levels, using the adaptive sampling mechanism induces a penalty in run time. This is due to the fact that processes need to synchronize to restart the computations, and the HSS tree has to be traversed multiple times instead of once. Also, instead of being computed in one shot, the samples \(S^r=A R^r\) and \(S^c=A^*R^c\) are computed in multiples stages, which mitigates the benefits of BLAS3 kernels. However, when the tree has more levels, we can see that the adaptive strategy can be faster than using directly the correct number of random vectors (3,000 here). This is due to the fact that, at the bottom of the tree, nodes have ranks much lower than 3,000. Their generators can be computed with less random vectors (e.g., 500 or 1,000). Using less random vectors makes the Interpolative Decomposition faster, and it can potentially make the whole compression stage faster. 

\begin{table}[!ht]\centering
\newcommand{\ce}[1]{\multicolumn{1}{c|}{#1}}%
\newcommand{\ece}[1]{\multicolumn{1}{|c|}{#1}}%
\setlength{\tabcolsep}{1.5em}%
\hspace{3.5em}\begin{tabular}{|r|r|r|r|}
\hline
Levels     & \multicolumn{3}{c|}{Strategy} \\
\cline{2-4}
\ece{in HSS}   & Steps of &       Steps of &       No adaptive \\
\ece{tree} & \ce{500} &     \ce{1,000} &       \ce{sampling} \\
\hline
     3 &    259.9 &          223.9 & \textbf{186.2} \\
\hline
     8 &    147.0 & \textbf{125.1} &          137.3 \\ 
\hline
\end{tabular}\hspace{3.5em}
\caption{Time in seconds for the HSS compression as a function of the number of levels in the HSS tree and the sampling strategy (as in Figure~\ref{fig:adaptive}). The matrix arises from the discretization of an electromagnetic sphere using BEM and has size 58,800.\label{tab:adaptive}}
\end{table}

In a practical setting, it is impossible to predict what the fastest strategy is. However, in many applications, practitioners have a rough idea of the compressibility of their matrices and can predict the order of magnitude of the maximum rank. In that case, we advise to set the sampling parameters so that, at worst, the compression routine needs to restart a limited number of times. For example, if the rank is expected to be between 1,000 and 10,000, we would start with 1,000 random vectors and increase the number by 1,000 every time a step of compression fails, guaranteeing no more than 10 steps.
 
\subsection{Scalability}
\label{sec:scal}

In this section, we evaluate the scalability of our structured code using three experiments.

In the first experiment, we process dense matrices with increasing size from the same application, using an increasing number of MPI processes. The experimental setting is the same as in~\cite{wang2013efficient}; we use the same system (Hopper at NERSC) and the same settings. The matrices we consider correspond to the root node of the multifrontal factorization of the discretized Helmholtz equations, with a fixed number of points per wavelength. They correspond to five different cubic meshes, ranging from \(100\times100\times100\) to \(500\times500\times500\). The topmost separators, i.e., the dense matrices we consider here, have between \(10,000\) and \(250,000\) rows and columns. In Table~\ref{tab:resRoot}, we compare the performance of ScaLAPACK, Hsolver (more precisely, the dense kernel used within Hsolver), and STRUMPACK when solving a linear system with these matrices. Under this setup, the maximum HSS rank grows linearly with the mesh size \(k\). Note that this is not strictly a weak scaling experiment since the number of processes does not increase as fast as the number of operations. The next experiment in this section is a strict weak scaling experiment.

\begin{table}[!ht]\centering
\renewcommand{\arraystretch}{1.25}
\setlength{\tabcolsep}{0.2em}
\newcommand{\ce}[1]{\multicolumn{1}{c|}{#1}}
\newcommand{\cee}[1]{\multicolumn{1}{c||}{#1}}
\begin{tabular}{|c|l||r||r||r||r||r|}
\hhline{|--||-||-||-||-||-|}
\multicolumn{2}{|c||}{\(k\) (mesh: \(k\times k\times k\))}                                                                          &    \cee{100} &    \cee{200} &    \cee{300} &    \cee{400} &     \ce{500} \\ 
\hhline{|--||-||-||-||-||-|}
\multicolumn{2}{|c||}{Matrix size (=\(k^2\))}                                                                                       &       10,000 &       40,000 &       90,000 &      160,000 &      250,000 \\ 
\hhline{|--||-||-||-||-||-|}
\multicolumn{2}{|c||}{MPI tasks}                                                                                                    &     \cee{64} &    \cee{256} &  \cee{1,024} &  \cee{4,096} &   \ce{8,192} \\ 
\hhline{==::=::=::=::=::=}
\multirow{3}{*}{\rotatebox{90}{\phantom{La}Sca}\ \rotatebox{90}{\scalebox{0.95}{LAPACK}}} &               Flops (\(\times10^{12}\)) &          2.7 &        170.7 &       1944.0 &      10922.7 &      41666.7 \\ 
\hhline{|~-||-||-||-||-||-|}
                                                                                          &                                Time (s) &          4.2 &         57.7 &        176.1 &        313.6 &        541.6 \\ 
\hhline{|~-||-||-||-||-||-|}
                                                                                          &                      Communication time &       30.5\% &       20.5\% &       24.9\% &       40.4\% &       36.1\% \\ 
\hhline{==::=::=::=::=::=}
\multirow{10}{*}{\rotatebox{90}{Hsolver}}                                                 &                            Maximum rank &          335 &          618 &          894 &         1226 &         1497 \\ 
\hhline{|~-||-||-||-||-||-|}
                                                                                          &                        HSS factors (GB) &          0.1 &          0.8 &          2.0 &          4.6 &          6.8 \\ 
\hhline{|~-||-||-||-||-||-|}
                                                                                          &   Compression flops (\(\times10^{12}\)) &          0.8 &         19.7 &        115.2 &        424.0 &       1051.0 \\ 
\hhline{|~-||-||-||-||-||-|}
                                                                                          &                    Compression time (s) &          8.3 &         51.5 &        193.4 &        207.8 &        259.5 \\ 
\hhline{|~-||-||-||-||-||-|}
                                                                                          & Factorization flops (\(\times10^{12}\)) &          0.1 &          0.7 &          2.3 &          7.2 &         10.6 \\ 
\hhline{|~-||-||-||-||-||-|}
                                                                                          &                  Factorization time (s) &          0.4 &          1.4 &          1.8 &          2.5 &          4.2 \\ 
\hhline{|~-||-||-||-||-||-|}
                                                                                          &       Solution flops (\(\times10^{9}\)) &          0.1 &          0.3 &          0.8 &          2.1 &          2.9 \\ 
\hhline{|~-||-||-||-||-||-|}
                                                                                          &                       Solution time (s) &          0.1 &          0.2 &          0.6 &          2.3 &          9.5 \\ 
\hhline{|~-||-||-||-||-||-|}
                                                                                          &                      Communication time &       12.4\% &       19.4\% &       27.7\% &       26.4\% &       31.3\% \\ 
\hhline{|~-||-||-||-||-||-|}
                                                                                          &                 Speed-up over ScaLAPACK &          0.5 &          1.1 &          0.9 &          1.5 &          2.0 \\ 
\hhline{==::=::=::=::=::=}
\multirow{11}{*}{\rotatebox{90}{STRUMPACK}}                                               &                            Maximum rank &          313 &          638 &          903 &         1289 &         1625 \\ 
\hhline{|~-||-||-||-||-||-|}
                                                                                          &                        HSS factors (GB) &          0.1 &          0.5 &          1.1 &          3.0 &          3.3 \\ 
\hhline{|~-||-||-||-||-||-|}
                                                                                          &   Compression flops (\(\times10^{12}\)) &          0.6 &         18.8 &        132.7 &        626.1 &       1716.7 \\ 
\hhline{|~-||-||-||-||-||-|}
                                                                                          &                    Compression time (s) &          2.0 &         13.0 &         30.6 &         60.8 &        133.6 \\ 
\hhline{|~-||-||-||-||-||-|}
                                                                                          & Factorization flops (\(\times10^{12}\)) &         0.04 &          0.5 &          1.7 &          5.9 &          7.7 \\ 
\hhline{|~-||-||-||-||-||-|}
                                                                                          &                  Factorization time (s) &          0.3 &          1.0 &          1.5 &          2.7 &          5.0 \\ 
\hhline{|~-||-||-||-||-||-|}
                                                                                          &       Solution flops (\(\times10^{9}\)) &          0.2 &          1.1 &          2.9 &          6.9 &          9.5 \\ 
\hhline{|~-||-||-||-||-||-|}
                                                                                          &                       Solution time (s) &         0.04 &          0.3 &          0.4 &          0.6 &          0.8 \\ 
\hhline{|~-||-||-||-||-||-|}
                                                                                          &                      Communication time &       24.2\% &       24.0\% &       27.0\% &       28.5\% &       28.9\% \\ 
\hhline{|~-||-||-||-||-||-|}
                                                                                          &                 Speed-up over ScaLAPACK & \textbf{1.8} & \textbf{4.0} & \textbf{5.4} & \textbf{4.8} & \textbf{3.9} \\ 
\hhline{|~-||-||-||-||-||-|}
                                                                                          &                   Speed-up over Hsolver & \textbf{3.8} & \textbf{3.7} & \textbf{6.0} & \textbf{3.3} & \textbf{2.0} \\ 
\hhline{|--||-||-||-||-||-|}
\end{tabular}
\caption{Comparison between ScaLAPACK, Hsolver, and STRUMPACK for dense matrices arising from the multifrontal factorization of the discretized Helmholtz equations.~\label{tab:resRoot}}
\end{table}

One can observe that STRUMPACK and Hsolver find similar maximum ranks for all the problems. The size of HSS factors is smaller with STRUMPACK, which is due to the special structure of \(U\) and \(V\) generators, as described in Section~\ref{sec:compr}. In terms of performance, STRUMPACK is 2 to 6 times faster than Hsolver, and 1.8 to 5.4 faster than ScaLAPACK. It is interesting to notice that STRUMPACK spends less time in communication. The percentage of wall time spent in communications, as reported by the IPM tool~\cite{fuerlinger2010effective}, is similar to that of Hsolver, but the overall wall time is shorter, implying less time is spent doing communication (assuming computations and communications do not overlap, which is fair since our algorithm is mostly synchronous). 

The second experiment in this section is a strict weak-scaling experiment. We consider the root node of the multifrontal factorization of the discretized Poisson equation on a 2D mesh. The mesh is a \(k\times k\) regular grid, therefore the dense matrix that we consider (last frontal matrix) is \(k\times k\). The multifrontal factorization yields frontal matrices that can be compressed using HSS techniques with a very low maximum rank~\cite{xia2013efficient}, that is almost constant with respect to the size of the grid (in practice, it increases very slowly -- logarithmically). In the experiment, we use a fixed number of random vectors (slightly larger than the rank of the largest problem). Therefore, the complexity of the HSS compression grows as \(O(k^2)\). In the experiment, the number of processes also grows as \(k^2\), yielding a constant number of operations per process for the different grid sizes, as shown in Table~\ref{tab:resLapl}. One can observe that the run time increases as \(k\) increases. This is due to the overhead of communications. In particular, for the last problem, the redistribution of the input matrix represents over 80\% of the compression time. This is what is expected for problems with very small maximum rank, as shown in Section~\ref{sec:comm}, Table~\ref{tab:comm_costs}. If we ignore the redistribution time, then compression time is reasonably constant, as shown in the last row of the table.

\begin{table}[!ht]\centering
\renewcommand{\arraystretch}{1.25}
\setlength{\tabcolsep}{0.2em}
\newcommand{\ce}[1]{\multicolumn{1}{c|}{#1}}
\newcommand{\cee}[1]{\multicolumn{1}{c||}{#1}}
\begin{tabular}{|l||r||r||r||r||r|}
\hhline{|-||-||-||-||-||-|}
\multicolumn{1}{|c||}{\(k\) (matrix size: \(k\times k\))} & \cee{2,000} & \cee{4,000} & \cee{8,000} & \cee{16,000} & \ce{32,000} \\ 
\hhline{|-||-||-||-||-||-|}                                             
\multicolumn{1}{|c||}{MPI tasks}                          &     \cee{1} &     \cee{4} &    \cee{16} &     \cee{64} &    \ce{256} \\ 
\hhline{=::=::=::=::=::=}                                               
Maximum rank                                              &          21 &          26 &          31 &           38 &          44 \\ 
\hhline{|-||-||-||-||-||-|}                                             
Compression flops per process (\(\times10^{9}\))          &        1.08 &        1.06 &        1.06 &         1.06 &        1.05 \\ 
\hhline{|-||-||-||-||-||-|}                                             
Compression time (s)                                      &        0.10 &        0.21 &        0.37 &         0.54 &        2.51 \\ 
\hhline{|-||-||-||-||-||-|}
Compression time w/o redistribution (s)                   &        0.10 &        0.11 &        0.13 &         0.20 &        0.42 \\
\hhline{|-||-||-||-||-||-|}
\end{tabular}
\caption{Weak-scaling experiment for dense matrices arising from the multifrontal factorization of the discretized 2D Poisson equation.~\label{tab:resLapl}}
\end{table}

The last experiment in this section is a strong scaling benchmark. We use one test problem, a matrix arising from the discretization of an electromagnetic sphere using BEM, with size 130,000. We compare the run time for solving a linear system with ScaLAPACK and STRUMPACK, using a number of MPI processes ranging from 256 to 4,096. For this problem, the maximum rank is 5,500. 

\begin{figure}[!ht]
\pgfplotstableread{
{MPI} Ideal ScaLA Strum
  256 862.4 862.4 469.7
  512 431.2 466.8 233.6
 1024 215.6 257.4 176.3
 2048 107.8 157.7 130.2
 4096  53.9 132.9 121.9
}{\myresults}\hspace{-5pt}%
\subfigure[Run time.\label{fig:strongScalingPlot}]{
\begin{tikzpicture}[scale=.86,font=\large]
\begin{loglogaxis}[xlabel=MPI tasks, ylabel=Run time, xtick=data, xticklabels from table={\myresults}{MPI}, ylabel near ticks]
\addplot[color=black,dashed] table[x index=0,y index=1]{\myresults};
\addplot[color=red,mark=square*,dotted,very thick, mark options={solid,scale=1.3}] table[x index=0,y index=2]{\myresults};
\addplot[color=blue,mark=*,dotted,very thick,mark options={solid,scale=1.4}] table[x index=0,y index=3]{\myresults};
\legend{Ideal scaling,ScaLAPACK,STRUMPACK}
\end{loglogaxis}
\end{tikzpicture}\ 
}\hfill
\subfigure[Statistics.\label{fig:strongScalingData}]{\raisebox{55pt}{\small
\renewcommand{\arraystretch}{1.25}
\setlength{\tabcolsep}{0.1em}
\newcommand{\ce}[1]{\multicolumn{1}{c|}{#1}}
\newcommand{\cee}[1]{\multicolumn{1}{c||}{#1}}
\begin{tabular}{|c|l||r||r||r||r||r|}
\hhline{|--||-||-||-||-||-|}
\multicolumn{2}{|c||}{MPI tasks} & \cee{256} & \cee{512} & \cee{1,024} & \cee{2,048} & \ce{4,096} \\ 
\hhline{==::=::=::=::=::=}
\multirow{2}{*}{LU}   & Time (s) &     862.4 &     466.8 &       257.4 &       157.7 &      132.9 \\
\hhline{|~-||-||-||-||-||-|}
                      & \% comm. &    30.5\% &    37.9\% &      45.0\% &      55.9\% &     74.3\% \\ 
\hhline{==::=::=::=::=::=}
\multirow{2}{*}{HSS}  & Time (s) &     469.7 &     233.6 &       176.3 &       130.2 &     121.9  \\ 
\hhline{|~-||-||-||-||-||-|}
                      & \% comm. &    34.7\% &    31.2\% &      42.1\% &      47.3\% &     68.0\% \\ 
\hhline{|--||-||-||-||-||-|}
\end{tabular}
}}
\caption{Strong scaling experiment: run time for solving a linear system for a 
matrix with size 130,000, arising from the discretization of an electromagnetic
 sphere using BEM, with maximum rank 5,500.}
\label{fig:strongScaling}
\end{figure}

We observe that although the scalability of STRUMPACK is quite good, the gap between ScaLAPACK and STRUMPACK reduces when the number of processes increases. As explained in Section~\ref{sec:comm}, this is because when the HSS rank is large (which is the case in this problem), communication volume for the traversal of the tree becomes larger than that with ScaLAPACK. The breakdown of the run time for the parallel HSS compression with 4,096 MPI tasks is the following: 15\% of the time is spent in the initial matrix distribution, and 25\% of the time is spent in the two products \(S^r=A R^r\) and \(S^c=A^* R^c\). The rest (60\%) is spent traversing the HSS tree to compute the local samples and generators. The major part is spent computing Interpolative Decompositions, which represent 50\% of total compression time. We observed that in most cases the flop-rate of the Interpolative Decomposition (modified version of \texttt{PxGEQPF}) is much lower than that of \texttt{PxGEMM} or \texttt{PxGETRF}. This is due to the fact that it relies on a BLAS2 algorithm. A BLAS3 implementation appears in the literature but the code is not publicly available~\cite{benner2005parallel}. Implementing a BLAS3 version is left for future work.

We are investigating different techniques to accelerate the distribution of the input matrix \(A\). Furthermore, some recent works investigate communication optimal matrix-matrix multiplication algorithms~\cite{solomonik2011communication} and improvements for rectangular matrix multiplications~\cite{demmel2013communication}. Our implementation would directly benefit from any improvement resulting from this research.

\section{Conclusion}

We presented the dense matrix computation package STRUMPACK that uses Hierarchically Semi-Separable representations to compress an input matrix and performs operations with this compressed form, such as solving linear systems or performing matrix-vector products. For matrices from certain classes of applications, such as finite element or boundary element methods, or applications that involve Toeplitz matrices, using HSS techniques allows to perform these operations asymptotically faster than when traditional algorithms (e.g., \(LU\) factorization) are used. The compression algorithm, which is the cornerstone of the framework, is parametrized by a compression threshold that allows the package to be used as a direct solver with full accuracy or as a robust preconditioner. Our compression algorithm employs randomized sampling and is the first distributed-memory implementation that we know of. Furthermore, we introduced an adaptive sampling mechanism that allows the code to be used in a black-box fashion.

The STRUMPACK package is very general; it can be used with any number of MPI processes and can accommodate different hierarchical partitionings of the input matrix. Furthermore, it is an open source package made available to the community. The code is released under the BSD-LBNL license and the version presented here is currently available at \url{http://portal.nersc.gov/project/sparse/strumpack/STRUMPACK-Dense-0.9.0.tar.gz}.

Work is in progress to use this dense package within a sparse solver. We have also developed a shared-memory sparse solver~\cite{ghysels2014multifrontal} and our goal is to combine these two codes in order to obtain a hybrid (MPI+OpenMP) sparse solver. Another aspect that we wish to explore is using HSS techniques in matrix-free frameworks. As mentioned here, our algorithm is amenable to a matrix-free implementation where the user only provides a matrix-vector product and a routine to access selected elements of the matrix on the fly. This feature will be included in a future version of STRUMPACK.

\paragraph*{Acknowledgments}
Partial support for this work was provided through Scientific Discovery through Advanced Computing (SciDAC) program funded by U.S. Department of Energy, Office of Science, Advanced Scientific Computing Research (and Basic Energy Sciences/Biological and Environmental Research/High Energy Physics/Fusion Energy Sciences/Nuclear Physics).
This research used resources of the National Energy Research Scientific Computing Center, which is supported by the Office of Science of the U.S. Department of Energy under Contract No. DE-AC02-05CH11231.

We wish to thank people who provided us with test problems and helped us: Ana Manic, Guillaume Sylvand, Umberto Villa.

\bibliographystyle{plain}
\bibliography{paper}

\newpage

\appendix

\section{Two-stage triangular solution process}
\label{app:solve}

This appendix illustrates the ULV solve algorithm~\ref{alg:solve} starting from expression~\eqref{eqn:2level_ulv_transform}, giving the explicit ULV factorization of \(A\) for a 3 level HSS matrix. After ULV factorization, the solution of \(Ax=b\) can be obtained as \(x = V^{-1} L^{-1} U^{-1}b\).  In~\eqref{eqn:2level_ulv_transform}, the transformations applied to \(A\) from the left form \(U^{-1}\), the transformations applied to \(A\) from the right form \(V^{-1}\) and the big matrix in the right-hand side of Equation~\eqref{eqn:2level_ulv_transform} forms \(L\). Define
\begin{equation}
  \tilde{V}_{\tau} = Q_{\tau} \hat{V}_\tau \quad \text{with} \quad
  \hat{V}_{\tau} =
  \begin{cases}
    V_{\tau}, & \text{if \(\tau\) is a leaf} \\
    \begin{bmatrix}
      \tilde{V}_{\nu_1;b} & \\ & \tilde{V}_{\nu_2;b}
    \end{bmatrix} V_{\tau}, & \text{if \(\tau\) is a non-leaf}
  \end{cases} \\
\end{equation}
Let \(b_{\tau} = b(I_\tau)\) for leaves \(\tau\) and
\(\tilde{b}_{\tau} = \Omega_\tau b_{\tau}\).  Now, we first compute
\(U^{-1}b\) by applying the \(\Omega_\tau\) transformations and the
permutations \(\Gamma_{\nu_1;b\leftrightarrow \nu_2;t}\) to the
right-hand side \(b\)
\begin{align}
  U^{-1}b &= \Gamma_{1;b\leftrightarrow 2;t}
  \begin{bmatrix} I & & & \\ & \Omega_1 & &  \\  & & I & \\ &&& \Omega_2 \end{bmatrix}
  \begin{bmatrix} \Gamma_{3;b\leftrightarrow 4;t}\! & \\ & \!\Gamma_{5;b\leftrightarrow 6;t}\end{bmatrix}
  \begin{bmatrix} \Omega_3 b_3 \\ \Omega_4 b_4 \\ \Omega_5 b_5 \\ \Omega_6 b_6 \end{bmatrix} \\
  &= \Gamma_{1;b\leftrightarrow 2;t}
  \begin{bmatrix} I & & & \\ & \Omega_1 & &  \\  & & I & \\ &&& \Omega_2 \end{bmatrix}
  \begin{bmatrix} \tilde{b}_{3;t} \\ \tilde{b}_{4;t} \\ \tilde{b}_{3;b} \\ \tilde{b}_{4;b} \\
    \tilde{b}_{5;t} \\ \tilde{b}_{6;t} \\ \tilde{b}_{5;b} \\ \tilde{b}_{6;b} \end{bmatrix}
  = \Gamma_{1;b\leftrightarrow 2;t}
  \begin{bmatrix} \tilde{b}_{3;t} \\ \tilde{b}_{4;t} \\ \Omega_1 \begin{bmatrix} \tilde{b}_{3;b} \\ \tilde{b}_{4;b} \end{bmatrix} \\
    \tilde{b}_{5;t} \\ \tilde{b}_{6;t} \\ \Omega_2 \begin{bmatrix} \tilde{b}_{5;b} \\ \tilde{b}_{6;b} \end{bmatrix} \end{bmatrix}
  = \begin{bmatrix} \tilde{b}_{3;t} \\ \tilde{b}_{4;t} \\ \Omega_{1;t} \begin{bmatrix} \tilde{b}_{3;b} \\ \tilde{b}_{4;b} \end{bmatrix} \\
    \tilde{b}_{5;t} \\ \tilde{b}_{6;t} \\ \Omega_{2;t} \begin{bmatrix} \tilde{b}_{5;b} \\ \tilde{b}_{6;b} \end{bmatrix} \\
    \Omega_{1;b} \begin{bmatrix} \tilde{b}_{3;b} \\ \tilde{b}_{4;b} \end{bmatrix}  \\
    \Omega_{2;b} \begin{bmatrix} \tilde{b}_{5;b} \\ \tilde{b}_{6;b} \end{bmatrix} \end{bmatrix}
\end{align}
We write down explicitly the forward triangular substitution \(y=L^{-1}U^{-1}b\)
\begin{equation}
  y  = \begin{bmatrix}
    y_3 = L_3^{-1} \tilde{b}_{3;t} \\
    y_4 = L_4^{-1} \tilde{b}_{4;t} \\
    y_1 = L_1^{-1} \Omega_{1;t} \left( \begin{bmatrix} \tilde{b}_{3;b} \\ \tilde{b}_{4;b} \end{bmatrix} - L_{4,3} y_3 - L_{3,4} y_4 \right) \\
    y_5 = L_5^{-1} \tilde{b}_{5;t} \\
    y_6 = L_6^{-1} \tilde{b}_{6;t} \\
    y_2 = L_2^{-1} \Omega_{2;t} \left( \begin{bmatrix} \tilde{b}_{5;b} \\ \tilde{b}_{6;b} \end{bmatrix} - L_{6,5} y_5 - L_{5,6} y_6 \right) \\
    y_0 = D_0^{-1} \begin{bmatrix}
      \Omega_{1;b} \left( \begin{bmatrix} \tilde{b}_{3;b} \\ \tilde{b}_{4;b} \end{bmatrix} - L_{4,3} y_3 - L_{3,4} y_4 \right) -W_{1;b}Q_{1;t}^*y_1 - B_{1,2} V_2^*
      \begin{bmatrix} \tilde{V}_{5;t}^* & & \tilde{V}_{5;b}^* \\ & \tilde{V}_{6;t}^* & & \tilde{V}_{6;b}^*\end{bmatrix}
      \begin{bmatrix} y_5 \\ y_6 \\ Q_2^* y_2 \end{bmatrix} \\
      \Omega_{2;b} \left( \begin{bmatrix} \tilde{b}_{5;b} \\ \tilde{b}_{6;b} \end{bmatrix} - L_{6,5} y_5 - L_{5,6} y_6 \right) -W_{2;b}Q_{2;t}^*y_2 - B_{2,1} V_1^*
      \begin{bmatrix} \tilde{V}_{3;t}^* & & \tilde{V}_{3;b}^* \\ & \tilde{V}_{4;t}^* & & \tilde{V}_{4;b}^* \end{bmatrix}
      \begin{bmatrix} y_3 \\ y_4 \\ Q_1^* y_1 \end{bmatrix}
    \end{bmatrix}
  \end{bmatrix}
\end{equation}
Clearly, this substitution should be performed bottom-up, i.e., first compute the leaves \(y_3\), \(y_4\) \(y_5\) and \(y_6\), then \(y_1\) and \(y_2\) and finally \(y_0\). Now we introduce the intermediate variable \(z_\tau\), defined as
\begin{equation} \label{eq:definition_z}
  z_\tau = \begin{cases}
    \tilde{V}_{\tau;t}^* y_{\tau}, & \text{if \(\tau\) is a leaf} \\
    V_\tau^* \begin{bmatrix} z_{\nu_1} \\ z_{\nu_2} \end{bmatrix} + \tilde{V}_{\tau;t}^* y_\tau, & \text{if \(\tau\) is a non-leaf}
  \end{cases}
\end{equation}
Then for a non-leaf node \(\tau\) (f.i., nodes \(1\) and \(2\)), with two children \(\nu_1\) and \(\nu_2\) \emph{which are both leaves}, we have \(y_\tau = L^{-1}_{\tau} \Omega_{\tau;t} b_{\tau}\) with
\begin{align}
  b_{\tau} &= \left( \begin{bmatrix} \tilde{b}_{\nu_1;b} \\ \tilde{b}_{\nu_2;b} \end{bmatrix} - L_{\nu_2,\nu_1} y_{\nu_1} - L_{\nu_1,\nu_2} y_{\nu_2} \right) =
  \begin{bmatrix}
    \tilde{b}_{\nu_1;b} - W_{\nu_1;b}Q_{\nu_1;t}^* y_{\nu_1} - B_{\nu_1,\nu_2}V_{\nu_2}^*Q_{\nu_2;t}^* y_{\nu_2} \\
    \tilde{b}_{\nu_2;b} - B_{\nu_2,\nu_1}V_{\nu_1}^*Q_{\nu_1;t}^* y_{\nu_1} - W_{\nu_2;b}Q_{\nu_2;t}^* y_{\nu_2}
  \end{bmatrix} \\
  \label{eq:definition_btau}
  &= \begin{bmatrix}
    \tilde{b}_{\nu_1;b} - W_{\nu_1;b}Q_{\nu_1;t}^* y_{\nu_1} - B_{\nu_1,\nu_2} z_{\nu_2} \\
    \tilde{b}_{\nu_2;b} - B_{\nu_2,\nu_1} z_{\nu_1} - W_{\nu_2;b}Q_{\nu_2;t}^* y_{\nu_2}
  \end{bmatrix}
\end{align}
Due to the definition of \(z_\tau\) as given in~\eqref{eq:definition_z}, the definition of \(b_\tau\) for non-leaf nodes, Equation~\eqref{eq:definition_btau}, is also valid for nodes higher up in the hierarchy, for which the situation is slightly more complicated. Consider node \(0\), for which \(y_0 = D_0^{-1} b_0\), with
\begin{align}
  b_0  &= \begin{bmatrix}
    \tilde{b}_{1;b} - W_{1;b}Q_{1;t}^*y_1 - B_{1,2} \left( V_2^*
      \begin{bmatrix} \tilde{V}_{5;t}^* y_5 & \\ & \tilde{V}_{6;t}^* y_6 \end{bmatrix} + V_2^*
      \begin{bmatrix} \tilde{V}_{5;b}^* \\  & \tilde{V}_{6;b}^* \end{bmatrix} Q_2^* y_2 \right) \\
    \tilde{b}_{2;b} - W_{2;b}Q_{2;t}^*y_2 - B_{2,1} \left( V_1^*
      \begin{bmatrix} \tilde{V}_{3;t}^* y_3  \\ & \tilde{V}_{4;t}^*y_4 \end{bmatrix}  + V_1^*
      \begin{bmatrix} \tilde{V}_{3;b}^* \\ & \tilde{V}_{4;b}^* \end{bmatrix} Q_1^* y_1 \right)
  \end{bmatrix} \\
       &= \begin{bmatrix}
         \tilde{b}_{1;b} - W_{1;b}Q_{1;t}^* y_1 - B_{1,2} \left( V_2^*
           \begin{bmatrix} z_5 \\ z_6 \end{bmatrix} + \tilde{V}_{2;t}^* y_2 \right) \\
         \tilde{b}_{2;b} - W_{2;b}Q_{2;t}^*y_2 - B_{2,1} \left( V_1^*
           \begin{bmatrix} z_3 \\ z_4 \end{bmatrix} + \tilde{V}_{1;t}^* y_1 \right)
       \end{bmatrix} =
  \begin{bmatrix}
    \tilde{b}_{1;b} - W_{1;b}Q_{1;t}^* y_1 - B_{1,2} z_2 \\
    \tilde{b}_{2;b} - W_{2;b}Q_{2;t}^* y_2 - B_{2,1} z_1
  \end{bmatrix}
\end{align}
Hence, the \(z_\tau\) variables accumulate the contributions to the right-hand side from the already eliminated HSS nodes. We can compute \(y_\tau\) as \(y_\tau = L_{\tau}^{-1} \tilde{b}_\tau\), except at the root where \(y_0 = D_0^{-1} b_0\), which is computed using standard LU decomposition of \(D_0\). Finally, the orthogonal transformation \(V^{-1}\) involving the \(Q_\tau\) matrices should be applied to \(y\) to obtain the solution vector \(x\).

\end{document}